\documentclass[10pt,draftcls]{IEEEtran}
\def\journalname{Transaction of Control System Technology}
\usepackage{cite}
\usepackage{amsmath,amssymb,amsfonts}
\usepackage{algorithmic}
\usepackage{graphicx}
\usepackage{textcomp}
\usepackage{multirow}
\usepackage[utf8]{inputenc}
\usepackage[T1]{fontenc}
\usepackage{cuted}
\usepackage{multicol}
\usepackage{eurosym}
\usepackage{import}
\usepackage{datetime}
\usepackage{subcaption}
\usepackage{dblfloatfix}    

\usepackage[T3,T1]{fontenc}

\allowdisplaybreaks

\usepackage{xcolor}
\definecolor{REV}{RGB}{183, 10, 187}
\definecolor{REV2}{RGB}{183, 10, 87}

\graphicspath{ {Figures/} {figure/} {Figs/} {Matlab/}}

\newtheorem{assumption}{Assumption}
\newtheorem{remark}{Remark}

\usepackage[english]{babel}

\def\BibTeX{{\rm B\kern-.05em{\sc i\kern-.025em b}\kern-.08em
    T\kern-.1667em\lower.7ex\hbox{E}\kern-.125emX}}
\begin{document}
\title{A Hierarchical Architecture for Optimal Unit Commitment and Control of an Ensemble of Steam Generators}
\author{S.~Spinelli, M.~Farina, and A.~Ballarino
\thanks{Manuscript received on 03 July 2020. Revised: 22 March 2021. Accepted: 27 June 2021.
	 \textit{(Corresponding author: 	S.~Spinelli.) }}
\thanks{S.~Spinelli and A.~Ballarino are with Istituto di Sistemi e Tecnologie Industriali Intelligenti per il Manifatturiero 		Avanzato, Consiglio Nazionale delle Ricerche, Milano, Italy (e-mail:
	name.surname@stiima.cnr.it).	 }
\thanks{S.~Spinelli and M.~Farina are with Dipartimento di Elettronica, Informazione e Bioingegneria, Politecnico di Milano, Milano, Italy (e-mail: name.surname@polimi.it).}}

\maketitle

\begin{abstract}
A hierarchical architecture for the optimal management of an ensemble of steam generators is presented. The  subsystems are coordinated by a multi-layer scheme for jointly sustaining  a common load. The high level optimizes the load allocation and the generator schedule, considering activation dynamics by a hybrid model.
At medium level, a robust tube-based Model Predictive Control (MPC) tracks a time-varying demand using a centralized - but aggregate -  model, whose order does not scale with the number of subsystems. A nonlinear optimization, at medium level, addresses MPC infeasibility due to abrupt changes of ensemble configuration.
Low-level decentralized controllers stabilize the generators. This control scheme enables  the dynamical modification of the ensemble configuration and plug and play operations. Simulations demonstrate the approach potentialities.
\end{abstract}


\section{Introduction and problem statement}
\label{sec:intro}
Steam is  widely used in industrial processes, playing a primary role in production. 
In industrial applications requiring a large and possibly time-varying steam demand, a flexible and efficient generation solution is mandatory. Since boiler operation close to the lower generation limit is largely inefficient, in high-fluctuating demand scenarios the production efficiency can be very unsatisfactory.
In these cases, a virtual generation plant constituted of a set of smaller units working in parallel can be a viable alternative to operate a single boiler on a larger range \cite{tipsminimize}. A set of cooperative smaller units can be reconfigured to produce what demanded, enabling a quick and optimal connection/disconnection of subsystems, and considering the current and/or forecasted demand.\\ 
In line with this vision, the main objective of this work is the proposal of a hierarchical control scheme for the optimal unit commitment (UC) and management of a group of steam generators that work in a parallel configuration to sustain a cumulative steam demand.
\subsection{State of the art}
The coordination of independent (or interdependent) subsystems towards a main target characterizes different industrial applications, e.g.,  smart grids and electrical generation systems \cite{yang2013parallel}, thermal energy grids \cite{theo2016milp}, building heating and cooling systems \cite{conte2016optimal}, and distribution networks of steam, water, or compressed air \cite{paparella2013load}.
These complex plants share similar features: several (homogeneous) systems work in parallel to commonly supply an overall demand; each subsystem and the whole plant operate in a constrained range and the subsystems must cooperate in a scenario of limited shared resources.
The studies referred above focus on the optimal load sharing among the parallel systems. Actually, two main aspects must be addressed in this context: (i) the unit commitment and economic dispatch of the subsystems; 
(ii) the dynamic control of the overall plant and of the single subsystems.\\
The two problems are characterized by different time-scales and are commonly addressed separately. The UC optimization problem has been extensively studied in the context of electrical generation systems, where the scheduling is optimized to minimize the plant operating cost, while satisfying process (and market) constraints. Several approaches have been studied, both in the deterministic and stochastic framework: an extensive discussion about solution techniques can be found in the review papers \cite{saravanan2013solution,zheng2014stochastic}. While several (meta-)~heuristic methods and mathematical programming approaches have been tested in the literature, in this paper we address the solution of UC optimization by Mixed-Integer Programming (MIP), as it guarantees an efficient, flexible and accurate modeling framework.\\ In the context of combined cycle power plants, a MIP formulation for the scheduling of thermal units has been presented in \cite{arroyo2000optimal}, while a tighter formulation reducing the number of binary variables is presented in \cite{carrion2006computationally}. An extended formulation, that provides a generalized-mode model for each unit, is discussed in \cite{mitra2013optimal}. 
Discrete-time state-space model formulations can be easily implemented in MPC strategy to manage the plant in a receding horizon way, as discussed in \cite{tuffaha2017discrete}, whose formulation  permits only to describe the unit dynamics by ON/OFF modes. The one presented in \cite{ferrari2004modeling}, based on a hybrid system approach - and specifically on a Mixed-Logical Dynamical (MLD) model - can generalize the unit dynamics. Based on a similar approach, in \cite{spinelli2018hierarchical}, the authors have formulated  the high-level UC problem for a small  Combined Heat and Power (CHP)  unit, composed of a fire-tube boiler and an internal combustion engine for power generation.\\
In \cite{kopanos2018efficient}, the UC problem is presented for a CHP plant with eight steam boilers working in parallel, where maintenance issues of the flexible boiler array are integrated in the cost function.  The authors of \cite{dunn2009optimal} focus on the boiler load allocation problem, uncoupled from  electricity generation aspects, in a multi-boiler configuration:  the optimization is addresses by gradient search methods considering  boiler efficiency versus steam load. 
Crucially, these works focus only on the solution of the scheduling problem and do not consider the dynamic control of these units.\\
On the opposite front, other researchers are concentrating on dynamic control issues, with particular application on networked steam boilers operating in parallel. In \cite{bujak2009optimal} an optimal control scheme is presented for the energy loss minimization and the primary management of heat production for multi-boiler industrial systems, comparing the optimal approach to the traditional cascade control. The control of a multiple boiler configuration based on a MPC is discussed in  \cite{austin2010optimisation}, with application to a paper mill plant,  or in \cite{rambalee2010control} for a coal-fired boiler house, where maintaining stable header pressure and boiler availability is of critical importance for downstream consumers.\\
In the research work \cite{costanza2015optimal}, a supervisory control, designed by LQR approach, is studied for a set of boilers in parallel configuration: a dynamic feedback strategy allows to continuously change each boiler set-point, while minimizing a combined cost. Taking into account the dynamics of all the individual boilers, this optimal control can cope with general disturbances.  
However, the dimension of the model of the group of boilers grows with the number of units, thus encountering scalability issues. Moreover, the scheme is not flexible to dynamically manage the variation of the boiler number, i.e., enabling plug-and-play capabilities.\\
In the recent years, some efforts have been devoted to provide unitary solutions to these problems. In this respect, decentralized, distributed, and hierarchical methods have many advantages over centralized ones, in view of their flexibility, robustness (e.g., to system changes and demand variations), and scalability. In this work we focus on hierarchical methods,  as the elective choice for optimal supervision and coordination of the system ensembles, e.g., as introduced in \cite{Petzke2018}.\\
An extensive review of hierarchical and distributed approaches is reported in \cite{Scattolini2009}. 
Recently, different solutions have been proposed based on the receding horizon approach. For example, \cite{Bemporad2010} proposes a multi-rate solution for constrained linear systems based on reference governors, \cite{garone2017reference,kalabic2012reduced}; on the other hand, in \cite{Picasso2016} a hierarchical scheme is introduced for coordinating independent systems with joint constraints and \cite{Farina2018} extends the approach used in \cite{Picasso2016} in case of dynamically coupled units. Finally, \cite{Farina18} proposes a scalable solution based on finite impulse response models enabling plug-and-play operations, while \cite{sokoler2016hierarchical} presents an application on power systems. 

\emph{Notation}: 
	Calligraphic letters, $\mathcal{U, Y, W, Z}$, indicate sets. The Minkowski sum of two sets is denoted by $\oplus$, while $\bigoplus_{i=1}^{N_{\rm\scriptscriptstyle g}}\mathcal{W}_{\scriptscriptstyle i}=\mathcal{W}_{\scriptscriptstyle 1} \oplus \dots \oplus \mathcal{W}_{N_{\rm\scriptscriptstyle g}}$. Ensemble (resp. reference-model) variables are indicated with the notation $\bar{\cdot}$  (resp. $\hat{\cdot}$). Nonlinear models and linear counterparts are denoted by $\mathcal{S}$ and $\mathcal{L}$, respectively. 	
	Superscript $^{\rm\scriptscriptstyle CL}$ (resp. $^{\rm\scriptscriptstyle OL}$) connotes closed (resp. open) loop systems.
	Superscript $^{[{\rm\scriptscriptstyle M}]}$ (resp. $^{[{\rm\scriptscriptstyle H}]}$) denotes variables with sampling time $T_{\rm\scriptscriptstyle M}$ (resp. $T_{{\rm\scriptscriptstyle H}}$), whose discrete time index is $k$ (resp. $h$), referred to medium (resp. high) level.
	The floor operator is $\lfloor \cdot \rfloor$. Finally, for a generic variable $v(k)$, we denote $\Delta v(k)=v(k)-v(k-1)$.
\subsection{Problem statement and paper contribution}
In this work, we propose a hierarchical architecture for the management of an ensemble of steam generators. 
The aim is to manage a group of $N_{\rm\scriptscriptstyle g}$ steam generators, working in a parallel configuration to sustain a cumulative steam demand, $\bar{q}^{\rm\scriptscriptstyle Dem}_{\rm\scriptscriptstyle s}$. The objective is to guarantee the required steam flow rate, with the minimum operating cost. This implies both the minimization of fuel gas and the optimization of the network configuration (i.e., the partial contribution of each boiler to the overall demand), also considering the activation strategy.\\
The steam generator network is  assumed to be composed of \textit{similar} dynamical systems, i.e.,  having homogeneous quantities as inputs and outputs, but that might differ in physical dimensions, nominal production rate, consumption, and efficiency.\\
Each subsystem $i$ is a water-tube boiler: 
a pressurized water, denoted feed-water $q_{{\rm\scriptscriptstyle f},i}$, circulates inside the tube coil, forced to flow by a displacement pump, and it is heated by a natural gas burner, whose flow rate is $q_{{\rm\scriptscriptstyle g}, \scriptscriptstyle i}$. The heat, transmitted to the flowing fluid, induces a phase transition of the feed-water into steam. The steam flow rate generated is $q_{{\rm\scriptscriptstyle s},\scriptscriptstyle i}$. This design is characterized by extremely short start-up time and safe steam generation  with respect to the fire-tube boiler configuration, due to the limited volume of water.
The single subsystems and the network of generators are subject to input and output constraints. Both local and global variables are assumed to be defined in convex and compact sets, $\mathcal{U}_{\scriptscriptstyle i}$, $\mathcal{Y}_{\scriptscriptstyle i}$, $\bar{\mathcal{U}}$ and $\bar{\mathcal{Y}}$, i.e. 
\begin{subequations}
	\begin{align}
		q_{{\rm\scriptscriptstyle s} ,\scriptscriptstyle i} &\in \mathcal{U}_{\scriptscriptstyle i} & 
		q_{{\rm\scriptscriptstyle g} ,\scriptscriptstyle i} & \in \mathcal{Y}_{\scriptscriptstyle i } \\
		\bar{q}_{{\scriptscriptstyle\rm s} } &= \sum_{i=1}^{N_{\scriptscriptstyle\rm g}} q_{{\scriptscriptstyle\rm s},\scriptscriptstyle i} \in \bar{\mathcal{U}} &
		\bar{q}_{{\scriptscriptstyle\rm g} }& = \sum_{i=1}^{N_{\scriptscriptstyle\rm g}} q_{{\scriptscriptstyle\rm g},\scriptscriptstyle i} \in \bar{\mathcal{Y}} \label{eq:uy_all}
	\end{align}\label{eq:constraints_all} \end{subequations}%
\noindent The proposed hierarchical control scheme consists of three layers.\smallskip\\
The  \emph{high layer}  (HL) extends the preliminary solution, proposed by the authors in \cite{SpinelliIFAC} for a constant load demand, considering a time-varying demand and the discrete operating modes dynamics of the generators.
To this aim, the model of the high-level behavior of the system is here defined in detail in Section \ref{sec:Model-sub:HL_hyb}.
This model is exploited by the top layer to optimize the strategy, i.e., the generator schedule and the working conditions, in order to minimize the operating costs. The activation/inactivation of units must consider the high-level state of each units and the transition costs. 
This layer computes the optimal number of units active and the best shares of production to be allocated to each boiler based on the time-varying profile of the demand. 
With respect to \cite{Farina18} and \cite{SpinelliIFAC}, in this paper, the optimization program is reformulated on local steam flow rates, instead of directly optimizing the sharing factors, which avoids to introduce mixed-integer bilinear constraints.\smallskip\\
At the \emph{medium  layer} {(ML),  a robust MPC scheme is adopted, similarly to \cite{Petzke2018}.
This layer, considering the ensemble model, allows to  robustly track the overall demand.
The ensemble model is an aggregate low-order model of the network of active systems, defined in a scalable way.
Differently from \cite{Petzke2018}, in this work, we assume that the sharing factors can change during time. This condition must be opportunely handled by improving the formulation of the optimal control problem (OCP), in order to ensure at each time instant feasibility of the corresponding optimization program. We  propose a procedure - based on an alternative nonlinear MPC program - to drive the ensemble to the new configuration when a sharp transition is not feasible.\smallskip\\
At the \emph{lowest layer} (LL), a set of decentralized controllers is used. Proportional-integral (PI) regulators,  as currently used in industrial practice, stabilize the internal pressure to its set-point and track the individual requests.
 In this work, we opt for state-of-the-art regulators at low level, decoupled on pressure and flow-rate loops. This control layer exploits on purpose the embedded regulators, as provided by the generator producer, since the latter are actually neither open nor accessible for modifications, due to safety and regulatory issues. This choice permits to apply the proposed management architecture on brownfield, also on legacy systems. %

\section{The Boiler Models}
In this section we present the dynamical model of the high-pressure steam generators  used at the different layers.
For notational simplicity, the index $i$ will be dropped when clear from the context.
\subsection{Nonlinear physical model}
The continuous-time nonlinear dynamical model of the steam generator is derived from the drum-boiler model presented in \cite{aastrom2000drum}. Here the equations are adapted to the considered configuration: differently from drum-boilers, no accumulation exists in the water tubes and the drum is absent.
\begin{figure}[thb]
	\centering
	\includegraphics[width=.7\linewidth]{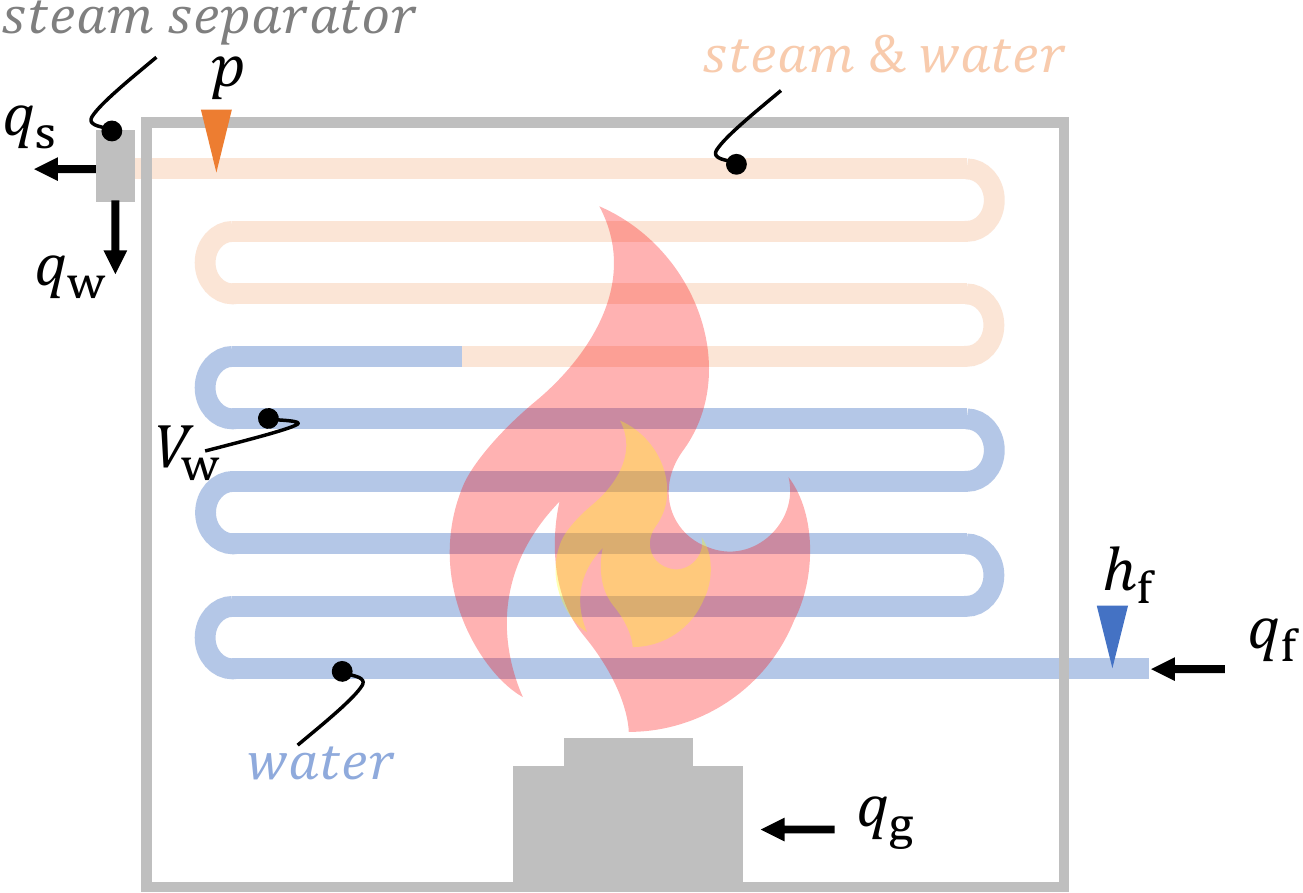}
	\caption{Steam generator functional scheme.}
	\label{fig:Boiler_scheme}
\end{figure}%
 In particular, the feed-water is forced to flow at high-pressure through the heated tubes, with flow-rate $q_{\scriptscriptstyle\rm f}$. The heat transfer transforms the feed-water either totally or partially  into steam. Therefore, the mass conservation equation on the water-tube control volume reads $ q_{\scriptscriptstyle\rm f} = q_{\scriptscriptstyle\rm s} + q_{\scriptscriptstyle\rm w}$, 
where $q_{\scriptscriptstyle\rm s}$ is the steam flow-rate. The portion of flow that persists in liquid phase at the outflow, $q_{\scriptscriptstyle\rm w}$,  is assumed to be at saturated temperature,  see Figure \ref{fig:Boiler_scheme}.\\
The $i$-th steam generator is characterized by a nonlinear dynamic model  $\mathcal{S}_{i}^{\scriptscriptstyle \rm{ OL}}$. 
\begin{align}
\dot{p} =  \frac{1}{\phi} (\eta \lambda_{\scriptscriptstyle H} q_{\scriptscriptstyle\rm g} + q_{\scriptscriptstyle\rm f}(h_{\scriptscriptstyle\rm f}-h_{\scriptscriptstyle\rm w}) - q_{\scriptscriptstyle\rm s}(h_{\scriptscriptstyle\rm s}-h_{\scriptscriptstyle\rm w})) \label{eq:nl_dyn1}\\
\dot{V}_{\scriptscriptstyle\rm w} =  \frac{1}{(\rho_{\scriptscriptstyle\rm w} - \rho_{\scriptscriptstyle\rm s})} (\frac{\partial\rho_{\scriptscriptstyle\rm w}}{\partial p}V_{\scriptscriptstyle\rm w} + \frac{\partial\rho_{\scriptscriptstyle\rm s}}{\partial p}V_{\scriptscriptstyle\rm s})\dot{p} \label{eq:nl_dyn2}
\end{align}
where
\begin{equation}
\resizebox{.9\hsize}{!}{$ \begin{split}\phi = V_{\scriptscriptstyle\rm s}(h_{\scriptscriptstyle\rm s}\frac{\partial\rho_{\scriptscriptstyle\rm s}}{\partial p} \!+\! \rho_{\scriptscriptstyle\rm s}\frac{\partial h_{\scriptscriptstyle\rm s}}{\partial p}) +V_{\scriptscriptstyle\rm w}(h_{\scriptscriptstyle\rm w}\frac{\partial\rho_{\scriptscriptstyle\rm w}}{\partial p} \!+\! \rho_{\scriptscriptstyle\rm w}\frac{\partial h_{\scriptscriptstyle\rm w}}{\partial p}) +\\ V_{\scriptscriptstyle\rm T} +  M_{\scriptscriptstyle\rm T} c_{\scriptscriptstyle\rm p} \frac{\partial T_{\scriptscriptstyle\rm s}}{\partial p} -(\frac{\partial\rho_{\scriptscriptstyle\rm w}}{\partial p}V_{\scriptscriptstyle\rm w} \!+\! \frac{\partial\rho_{\scriptscriptstyle\rm s}}{\partial p}V_{\scriptscriptstyle\rm s})\frac{(\rho_{\scriptscriptstyle\rm w}h_{\scriptscriptstyle\rm w} \!-\! \rho_{\scriptscriptstyle\rm s}h_{\scriptscriptstyle\rm s})}{(\rho_{\scriptscriptstyle\rm w} \!-\! \rho_{\scriptscriptstyle\rm s})}\end{split}$}\label{eq:phi}
\end{equation}
In equations \eqref{eq:nl_dyn1}-\eqref{eq:phi}, the subscripts $_{\rm{f,\,g,\,s,\,w}}$ refer to feed-water, fuel gas, steam, and internal water, respectively. 
Steam and internal water are assumed to be at saturated conditions. Therefore, the density $\rho$, the enthalpy $h$, and the temperature $T$ are only function of internal pressure $p$.\\ The system is characterized by some specific parameters: the burner efficiency $\eta$, the gas low heat value $\lambda_{\scriptscriptstyle H}$, the total tubes internal volume $V_{\scriptscriptstyle\rm T}$, the mass $M_{\scriptscriptstyle\rm T}$, and the specific heat coefficient $c_{\scriptscriptstyle\rm p}$.\\
The states of the nonlinear dynamical model \eqref{eq:nl_dyn1}-\eqref{eq:nl_dyn2} are the internal pressure $p$ and the water volume $V_{\scriptscriptstyle\rm w}$. The manipulable inputs are the feed-water flow rate $q_{\scriptscriptstyle\rm f}$ and the natural gas flow rate $q_{\scriptscriptstyle\rm g}$,
while the steam demand $q_{\scriptscriptstyle\rm s}$ is considered, at the low-level, as a disturbance term. Similarly, the enthalpy  $h_{\scriptscriptstyle\rm f}$ of the feed-water is considered a known measured disturbance.
\subsection{Low-level closed-loop model} \label{sec:low_ctrl}
An embedded controller is devoted to the regulation of the pressure at the set-point level, and to guarantee a constant water volume $V_{\scriptscriptstyle\rm w}$ for each subsystem  $\mathcal{S}_{i}^{\scriptscriptstyle \rm{OL}}$. This controller acts
on the local input variables  $q_{{\scriptscriptstyle\rm f}}$ and $q_{{\scriptscriptstyle\rm g}}$. 
Commercially-available boilers are already provided with low-level controllers for pressure regulation, designed on industrial standard configuration: a feedback PI regulator $\mathbf{R}$ on the fuel flow-rate, to steer the pressure $p$ to a set-point $p_{\rm sp}$, and a disturbance compensator $\mathbf{C}$  working, with an open-loop action, on the feed-water flow-rate to follow the steam demand, as depicted in Figure \ref{fig:SG_CL_reg}.
\begin{figure}[thb]
	\centering
	\includegraphics[width=1\linewidth]{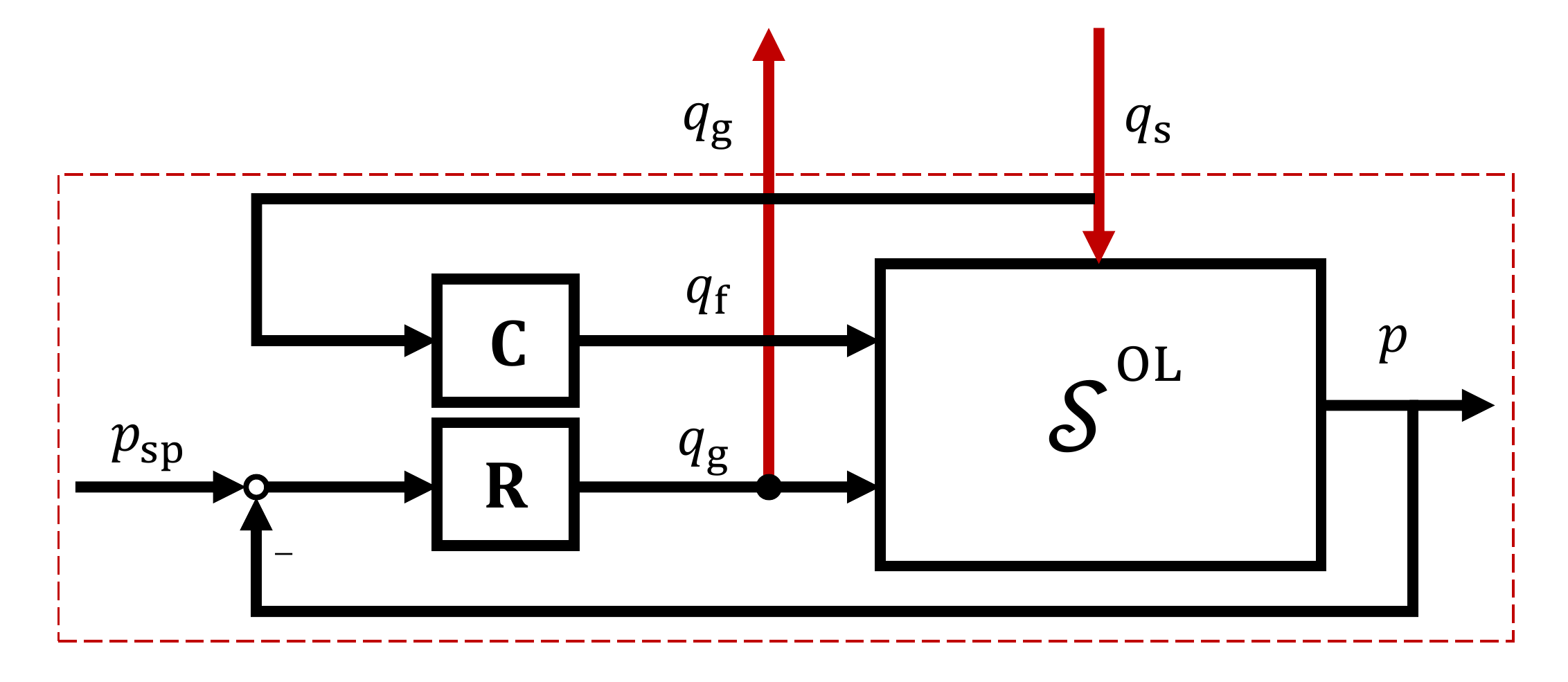}
	\caption{Closed-loop boiler function block diagram.}
	\label{fig:SG_CL_reg}
\end{figure}%
The closed-loop system of the $i$-th boiler can be described as a nonlinear dynamic model $\mathcal{S}_{i}^{\scriptscriptstyle \rm{CL}}$, in short denoted as  $ q_{{\rm g},i} = \mathcal{S}_{i}^{\scriptscriptstyle \rm{CL}}(q_{{\rm s},i})$.\\		 
One peculiarity of this closed-loop system is the possibility of considering the steam flow rate as input of the controlled system and the gas flow rate as an output, as shown in Figure \ref{fig:SG_CL_reg}. This closed-loop representation of the boiler enables the problem formalization  in the framework of hierarchical control of ensemble systems, as in \cite{Petzke2018}.\\
\begin{figure}[tbh]
	\centering
	\includegraphics[width=.9\linewidth]{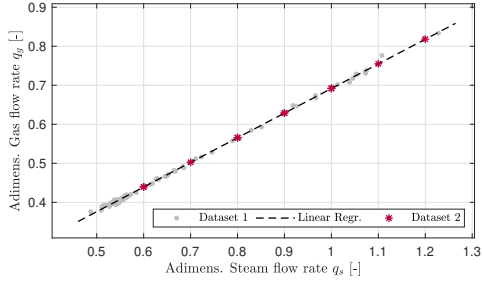}
	\caption{Input-output static map of the controlled steam generator at steady state.}
	\label{fig:AFFINE_static_map}
\end{figure}%
In Figure \ref{fig:AFFINE_static_map}, the input/output static map at steady state is shown: historical static data are compared with data generated simulating the response of the  system  $\mathcal{S}_{i}^{\scriptscriptstyle \rm{CL}}$ with a multiple step input profile. An affine approximation is also shown.  Note that, although this linear model is valid during production where the pressure is regulated at its set-point, non-linearity is still relevant during start-up.
\subsection{Affine model for medium-level control}
\label{sec:sub:L_CL}
Consistently  with the  data reported in Figure \ref{fig:AFFINE_static_map}, in production the boiler is maintained close to the nominal conditions, thus the dynamics of  $\mathcal{S}_{i}^{\scriptscriptstyle \rm{CL}}$ can be well represented by an affine dynamic model, used to account for transient response.
A discrete-time affine system  ${\mathcal{L}}_i^{\scriptscriptstyle \rm{CL}}$ with output $y(k)=q_{{\scriptscriptstyle\rm g},i}(kT_{\rm\scriptscriptstyle M})$ and input $u(k)=q_{{\scriptscriptstyle\rm s},i}(t)$ constant $\forall t\in[kT_{\rm\scriptscriptstyle M},(k+1)T_{\rm\scriptscriptstyle M})$
is identified with the simulation error minimization approach using the data drawn by exciting the controlled nonlinear model $\mathcal{S}_{i}^{\scriptscriptstyle \rm{ CL}}$ with multiple-step inputs. Note that the sampling time is $T_{\rm\scriptscriptstyle M}$ and the time index $k$ is the one used for control at medium hierarchical level. 
The identified discrete-time transfer function (plus constant) is denoted ${G}_{i}^{\scriptscriptstyle \rm{CL}}$ and is of the type
\begin{equation}
y(k)= \frac{\sum_{j=1}^{n_{\rm\scriptscriptstyle b}}(b_{j}z^{-j})}{1+ \sum_{j=1}^{n_{\rm\scriptscriptstyle f}}(f_{j}z^{-j})}u(k) + \gamma
\label{eq:poly_id}
\end{equation}
where $\gamma$ is the identified bias when $u(k)=0$. 
The corresponding state-space form is
\begin{equation}
\mathcal{L}_{i}^{\scriptscriptstyle \rm{CL}} : \left\lbrace  \begin{matrix}{{x}}(k+1) = {A} {x}(k) + {B} u(k) \\ y(k) =  {C} x(k) + {\gamma}\end{matrix} \right.
\label{eq:systems}
\end{equation}
with state vector, ${x}(k) =\linebreak[1] [\delta y({k}),\linebreak[1] ...,\linebreak[1] \delta y({k-n_{\scriptscriptstyle\rm f}}+1),\linebreak[1] {u}({k-1}),\linebreak[1] ...,\linebreak[1] {u}({k-n_{\scriptscriptstyle\rm b}}+1)]^{\rm\scriptscriptstyle  T}\in \mathbb{R}^{n_{\scriptscriptstyle\rm f}+n_{\scriptscriptstyle\rm b}-1}$ and $\delta y({k})=y(k)-\gamma$.  
The matrices are
$B=\begin{bmatrix}b_{\scriptscriptstyle 1}&0_{\scriptscriptstyle 1\times(n_{\scriptscriptstyle\rm f}-1)}&1&0_{\scriptscriptstyle 1\times(n_{\scriptscriptstyle\rm b}-2)}\end{bmatrix}^{\rm\scriptscriptstyle  T}$, $C=\begin{bmatrix}1&0&\dots&0\end{bmatrix}$, and
$$\arraycolsep=.5pt\def\arraystretch{1.1}
A\!=\!\left[\begin{array}{c|c}
\begin{array}{cc} -f_{\scriptscriptstyle 1} \dots  -f_{\scriptscriptstyle n_{\rm f}-2}& -f_{\scriptscriptstyle n_{\rm f}}\\I_{\scriptscriptstyle n_{\rm f}-1}&0_{\scriptscriptstyle (n_{\rm f}-1)\times 1}\end{array}
&\begin{array}{c}\begin{array}{lccr} b_{\scriptscriptstyle 2}  &\dots & b_{\scriptscriptstyle n_{\rm b}-1} & b_{\scriptscriptstyle n_{\rm b}}\end{array}\\
0_{\scriptscriptstyle (n_{\rm f}-1)\times(n_{\rm b}-1)}\end{array}\\\hline
\begin{array}{c}0_{\scriptscriptstyle (n_{\rm b}-1)\times n_{\rm f}}\end{array}&\begin{array}{cc}0_{\scriptscriptstyle 1\times (n_{\rm b}-2)}&0\\
I_{\scriptscriptstyle n_{\rm b}-2}&0_{\scriptscriptstyle (n_{\rm b}-2)\times 1}\end{array}
\end{array}\right]$$
\begin{assumption}
System~\eqref{eq:systems}, consistently with~\cite{Farina2018}, enjoys the following properties: 
\begin{enumerate}
	\item $A$ is Schur stable;
	\item $m=p=1$;
	\item $g=C (I_n - A)^{-1}B\neq 0$
\end{enumerate}
\label{ass:ass1}
\end{assumption}
This model will be used to derive the ensemble model, as discussed in detail in Section \ref{sec:M-MPC}. %
\subsection{Hybrid automaton for high-level optimization}
\label{sec:Model-sub:HL_hyb}
The boiler model used by the high-level optimizer operates on a coarser discrete-time grid, with a sampling time $T_{\rm\scriptscriptstyle H}$ and time index $h$.  
A hybrid automaton \cite{lygeros2003dynamical} is used, including both discrete and continuous states. The discrete variable $m$  defines the operating modes: shut down (OFF), start-up (ST), and production (ON), i.e., $m \in \{\rm{OFF, ST, ON}\}$.\\
A simplified model is considered in each mode, where the fuel flow rate $q_{\scriptscriptstyle\rm g}$ depends  on the steam demand $q_{\scriptscriptstyle\rm s}$. In this paper, due to the small settling time $t_{\scriptscriptstyle\rm st}$ of the dynamic system \eqref{eq:systems} with respect to the high-level sampling time, $t_{\scriptscriptstyle\rm st}\ll T_{\rm\scriptscriptstyle H}$,  a static input-output map for each operating mode is assumed.
Thus, the dynamic state of the Hybrid Automaton (DHA) is the number of sampling times $\chi$ spent in the present operation mode. Namely, variable $\chi\in \mathbb{Z}_0^+ $ is used for correctly model transitions.  A continuous evolution, $\mathsf{f}: \chi(h+1) = \chi(h)+1$ is valid when no transition occurs, i.e., $m(h+1)=m(h)$. Instead, any transition forces a reset to zero of the dynamic state, $\mathsf{r}:\chi(h+1) = 0$. 
	A mode transition - see the Finite State Machine (FSM) in Figure~\ref{fig:States} -
depends on the time spent in the current operating mode, $\chi(h)$, and possibly on switching binary input $\beta(h)\in\{0,1\}$  to $1$.
More specifically, a transition happens whenever a guard condition, $\mathsf{g}$, is met:
\begin{equation*}
\resizebox{1\hsize}{!}{$	\mathsf{g}:\left\lbrace  \begin{array}{ll}
		\{\chi(h)\geq \chi_{\scriptscriptstyle \rm{OFF}\rightarrow \rm{ST}}\}\wedge \{\beta(h)=1\} & m(h)=\rm{OFF}\\
		\{\chi(h)\geq \chi_{\scriptscriptstyle \rm{ST}\rightarrow \rm{ON}}\} & m(h)=\rm{ST}\\
		\{\chi(h)\geq \chi_{\scriptscriptstyle \rm{ON}\rightarrow \rm{OFF}}\}\wedge \{\beta(h)=1\} & m(h)=\rm{ON}\\
		\end{array}
	\right.$}
\end{equation*}
The values of $\chi_{\scriptscriptstyle \rm{OFF}\rightarrow \rm{ST}}$,  $\chi_{\scriptscriptstyle \rm{ST}\rightarrow \rm{ON}}$,  and $\chi_{\scriptscriptstyle \rm{ON}\rightarrow \rm{OFF}}$, are suitably-defined thresholds. The model output is given by:
 \begin{subequations}
 	\begin{align}
 	q_{\scriptscriptstyle\rm g}(h)&= g \cdot q_{\scriptscriptstyle\rm s}(h) + \gamma_{\scriptscriptstyle \rm ON} &\text{if }m&=\rm{ON}\label{eq:IN-OUT-B1}\\ 	
 	q_{\scriptscriptstyle\rm g}(h)&=  \gamma_{\scriptscriptstyle \rm ST}  &\text{if }m&=\rm{ST}\\ 	
 	q_{\scriptscriptstyle\rm g}(h)&=  0 &\text{if }m&=\rm{OFF}
 	\end{align}\label{eq:IN-OUT-B}\end{subequations}
where $g = C (I_n - A)^{-1}B $, $\gamma_{\scriptscriptstyle \rm ON}$, and $\gamma_{\scriptscriptstyle \rm ST}$ are the static gain of the closed-loop system  $\mathcal{L}^{\scriptscriptstyle \rm{CL}}$, the constant fuel gas consumption in production and in start-up modes, respectively. Note that, consistently with the model derived in the previous sections, the affine map \eqref{eq:IN-OUT-B1} is the one depicted in Figure \ref{fig:AFFINE_static_map}. 
\begin{figure}[bth]
	\centering

	\def\svgwidth{0.5\columnwidth}
	\includegraphics[width=.65\linewidth]{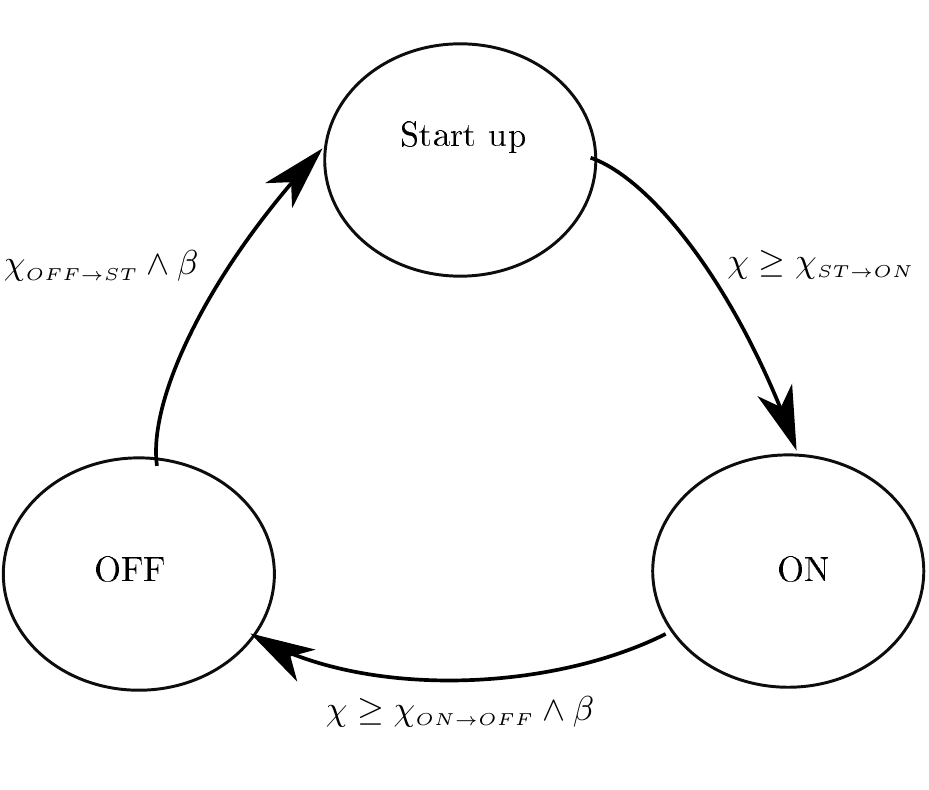}
	\caption{Boiler operation mode  transitions. }
	\label{fig:States}
\end{figure}
To make the model easily manageable in a suitable optimization program, the DHA model is converted into the  MLD one \cite{bemporad1999control}.
The MLD model is an extended state-space dynamical system where the state vector, $x^{\rm\scriptscriptstyle [H]}=\left\lbrace \chi, x^{\rm\scriptscriptstyle [H]}_{\scriptscriptstyle\rm OFF}, x^{\rm\scriptscriptstyle [H]}_{\scriptscriptstyle\rm ST}, x^{\rm\scriptscriptstyle [H]}_{\scriptscriptstyle\rm ON} \right\rbrace\in \mathbb{Z}\times\{0,1\}^3$, includes integer and Boolean variables. The inputs are the Boolean command and the steam flow-rate, $u^{\rm\scriptscriptstyle [H]}= \left\lbrace \beta^{\rm\scriptscriptstyle [H]}, q^{\rm\scriptscriptstyle [H]}_{\rm\scriptscriptstyle s}\right\rbrace\in \{0,1\}\times\mathbb{R}$, while the output is the consumed gas $y^{\rm\scriptscriptstyle [H]} = \left\lbrace  q^{\rm\scriptscriptstyle [H]}_{\rm\scriptscriptstyle g}\right\rbrace\in \mathbb{R}$, which depends on the active mode, as in \eqref{eq:IN-OUT-B}.\\
A set of Boolean and continuous auxiliary variables $\left\lbrace  \delta^{\rm\scriptscriptstyle [H]}, z^{\rm\scriptscriptstyle [H]} \right\rbrace\in\{0,1\}^{n_{\delta}}\times\mathbb{R}^{n_{z}}$ is added to model the FSM evolution, the transition guards, and the reset maps. The MLD model takes the general form: 
	\begin{equation*}
	\resizebox{1\hsize}{!}{$\begin{split}x^{\rm\scriptscriptstyle [H]}(h+1) =  A^{\rm\scriptscriptstyle [H]} x^{\rm\scriptscriptstyle [H]}(h) + B^{\rm\scriptscriptstyle [H]}_{u } u^{\rm\scriptscriptstyle [H]}(h) + B^{\rm\scriptscriptstyle [H]}_{z } z^{\rm\scriptscriptstyle [H]}(h) + B^{\rm\scriptscriptstyle [H]}_{\delta } \delta^{\rm\scriptscriptstyle [H]}(h)\\
y^{\rm\scriptscriptstyle [H]}(h) =  C^{\rm\scriptscriptstyle [H]} x^{\rm\scriptscriptstyle [H]}(h) + D_{u }^{\rm\scriptscriptstyle [H]} u^{\rm\scriptscriptstyle [H]}(h)+ D^{\rm\scriptscriptstyle [H]}_{z } z^{\rm\scriptscriptstyle [H]}(h) + D^{\rm\scriptscriptstyle [H]}_{\delta } \delta^{\rm\scriptscriptstyle [H]} (h) \\
E^{\rm\scriptscriptstyle [H]}_{x } x^{\rm\scriptscriptstyle [H]}(h) +  E^{\rm\scriptscriptstyle [H]}_{u } u^{\rm\scriptscriptstyle [H]}(h) +E^{\rm\scriptscriptstyle [H]}_{z } z^{\rm\scriptscriptstyle [H]}(h)  +E^{\rm\scriptscriptstyle [H]}_{\delta } \delta^{\rm\scriptscriptstyle [H]} (h)\leq E^{\rm\scriptscriptstyle [H]}_{\rm{aff}}\end{split}$}
	\end{equation*}%

\section{The hierarchical control scheme}
\label{sec:M_L-H_Lcontroller}
In the previous section we derived the single subsystem models to be used at the different levels.   
	Now, we explain how to manage and control them in a unitary and coordinated way.
\begin{figure}[hbt]
	\centering
	\includegraphics[width=1\linewidth]{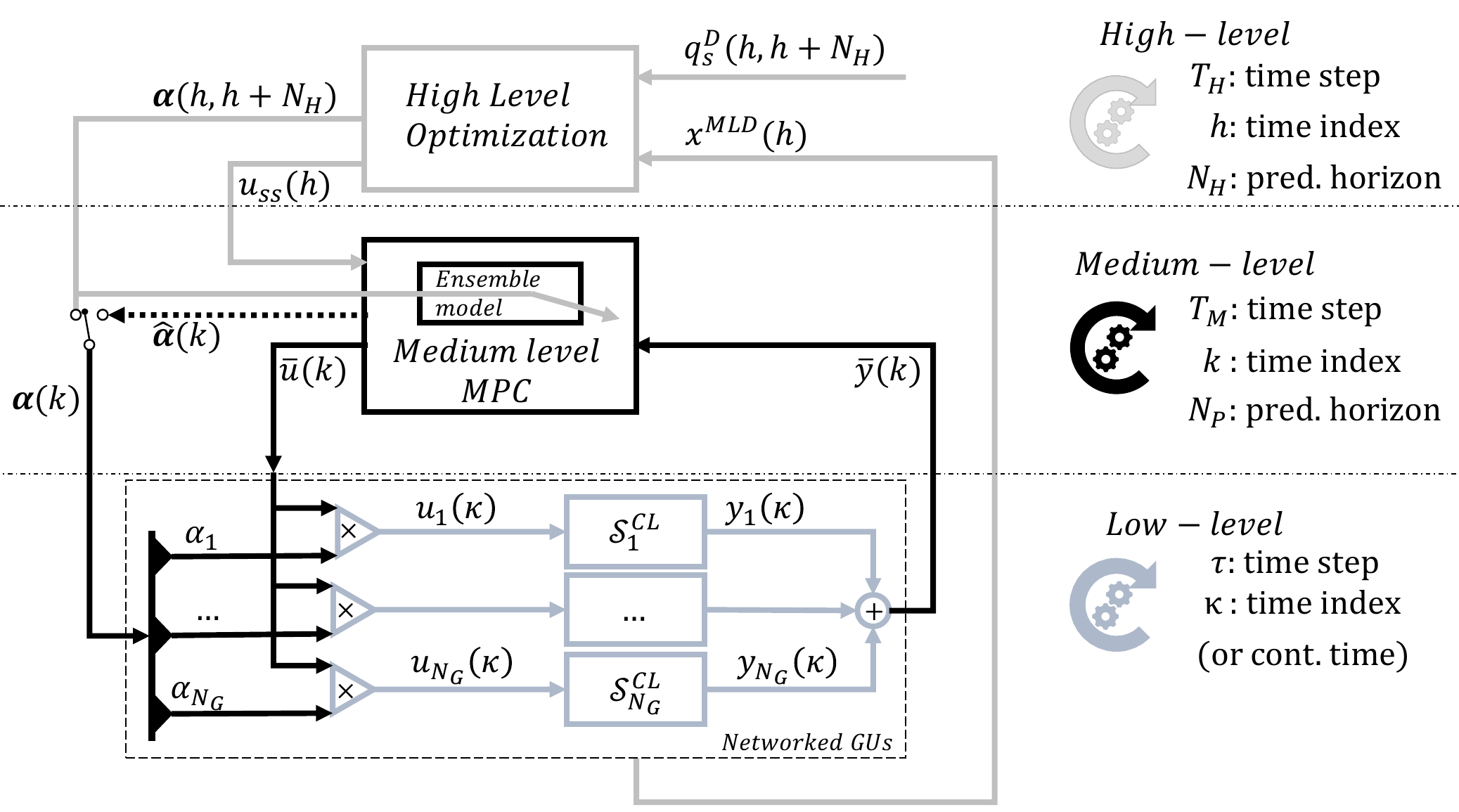}
	\caption{Steam generator ensemble and hierarchical scheme. Typically, $T_{\rm\scriptscriptstyle H}\in[10,30]$ min, $T_{\rm\scriptscriptstyle M}\in[30,60]$ s, $\tau\in[1,10]$ s.}
	\label{fig:SG_Ens}
\end{figure}%
\subsection{Sketch of the proposed control architecture}
As shown in Figure \ref{fig:SG_Ens}, the medium and high levels of the hierarchical scheme  are designed to concurrently define the input $u_i$ of each subsystem (i.e., local steam flow-rate $q_{{\rm\scriptscriptstyle s},i}$) as \begin{equation} u_{\scriptscriptstyle i} = \alpha_{\scriptscriptstyle i} \bar{u} \qquad i=1,\dots,N_{\rm\scriptscriptstyle g} 
	\end{equation}
where $\alpha_{\scriptscriptstyle i}$ is the sharing factor used to partition the overall ensemble input $\bar{u}$.\\
Sharing factors $\alpha_{\scriptscriptstyle i}$ are computed by the optimization layer. Here, thanks to the DHA models defined in Section~\ref{sec:Model-sub:HL_hyb}, they are optimized in a receding horizon way, minimizing the operating cost of the ensemble to supply the steam demand forecast. The sharing factors are time-varying and defined according to the slow time-scale (i.e., $T_{\rm\scriptscriptstyle H}$).\\
The ensemble input $\bar{u}$ is instead computed by a dynamic optimal reference tracking problem at medium level.
To do so, an aggregate model of the whole ensemble is derived by considering the subset of active generation units. The ensemble dynamical model is built by combining opportunely the closed-loop models of the controlled generators. By considering a unique ensemble model, medium level exhibits interesting scalability properties, as its dimensions do not grow with the number of subsystems.
A robust reference-tracking  MPC scheme is implemented to define the overall gas consumption of the ensemble, operating with a faster time-scale with respect to the high level, 	with sampling time $T_{\rm\scriptscriptstyle M}=T_{\rm\scriptscriptstyle H}$. %
\subsection{High-level Optimization} \label{sec:HL_opt}
The high hierarchical level aims to optimize the sharing factor profiles $\alpha^{\rm\scriptscriptstyle [H]}_i(h)$ and the modes of all the subsystems by minimizing the operating expenses, including the subsystem activation costs, the actual start-up time, and other constraints, as the ones related to mode transitions and operational range of each subsystem in the ensemble.\\
The algorithm presented here extends the one presented in~\cite{Petzke2018}, \cite{Farina18}, and {\cite{SpinelliIFAC}} by solving the unit commitment in receding horizon along a prediction window with time-varying demand.
We assume its profile to be known for the entire prediction horizon and approximated by a piece-wise constant function, $\bar{q}_{\scriptscriptstyle\rm s}^{\scriptscriptstyle\rm Dem}(h)$. \\
In {\cite{SpinelliIFAC}}, where both the sharing factors $\alpha^{\rm\scriptscriptstyle [H]}_{\scriptscriptstyle i}$ and the ensemble steady-state input $\bar{u}_{\scriptscriptstyle\rm ss}(h)$ were considered as decision variables, we obtained a MIP with bilinear inequality constraints. In this work the problem is reformulated as a simpler MIP with linear constraints by considering as optimization variables the partitioned steam flow rates $q^{\rm\scriptscriptstyle [H]}_{{\scriptscriptstyle\rm s}\, \scriptscriptstyle i}(h)$.
In this formulation, the optimal sharing factors are computed as 
$\alpha^{\rm\scriptscriptstyle [H]}_{\scriptscriptstyle i}(h) = {q^{\rm\scriptscriptstyle [H]}_{{\scriptscriptstyle\rm s}\,\scriptscriptstyle i}(h)}/{\bar{u}_{\scriptscriptstyle\rm ss}(h)}$.\\
The optimization problem at high-level reads:
\begin{subequations}
	\label{eq:opt_simpl}
	\begin{equation}
	\min_{\boldsymbol{\beta}^{\rm\scriptscriptstyle [H]}_{\scriptscriptstyle i}, \mathbf{q}^{\rm\scriptscriptstyle [H]}_{\scriptscriptstyle{\rm s}\, i}} \quad \textstyle\sum_{h=0}^{N_{\scriptscriptstyle\rm H}}\textstyle\sum_{i=1}^{N_{\rm\scriptscriptstyle g}} l_{\scriptscriptstyle i} (h, \beta_{\scriptscriptstyle i}(h), q^{\rm\scriptscriptstyle [H]}_{{\scriptscriptstyle\rm s}\,\scriptscriptstyle i}(h))
	\label{eq:alpha_opt_simpl}
	\end{equation}%
	\begin{align}
	\text{s.t.} \quad 
	& \textstyle\sum_{i=1}^{N_{\rm\scriptscriptstyle  g}} q^{\rm\scriptscriptstyle [H]}_{{\scriptscriptstyle\rm s}\, \scriptscriptstyle i}(h) \geq \bar{q}_{\scriptscriptstyle\rm s}^{\scriptscriptstyle\rm Dem}(h) \label{eq:qs_DemCon_simpl}\\
	& \left\lbrace \begin{array}{l}
			\textstyle\sum_{i=1}^{N_{\rm\scriptscriptstyle g}} q^{\rm\scriptscriptstyle [H]}_{{\scriptscriptstyle\rm s}\,\scriptscriptstyle i}(h) =0 		\\ \qquad
			\text{iff } \quad \textstyle\sum_{i=1}^{N_{\rm\scriptscriptstyle g}} x^{\rm\scriptscriptstyle [H]}_{\scriptscriptstyle{\rm{{ON}}} \,\scriptscriptstyle  i} (h)=0 	
			\\
			\bar{u}_{{\rm\scriptscriptstyle m}} \leq\textstyle\sum_{i=1}^{N_{\rm\scriptscriptstyle g}} q^{\rm\scriptscriptstyle [H]}_{{\scriptscriptstyle\rm s}\,\scriptscriptstyle i}(h) =0 \leq \bar{u}_{{\rm\scriptscriptstyle M}} 
			\\ \qquad 
			\text{otherwise }
	\end{array}\right.  \label{eq:u_glob}\\
	& \left\lbrace \begin{array}{l}
		0 \leq\textstyle\sum_{i=1}^{N_{\rm\scriptscriptstyle g}} q^{\rm\scriptscriptstyle [H]}_{{\scriptscriptstyle\rm g}\,\scriptscriptstyle i}(h)  \leq \bar{y}_{{\rm\scriptscriptstyle ST}} 		\\ \qquad
		\text{iff } \quad \textstyle\sum_{i=1}^{N_{\rm\scriptscriptstyle g}} x^{\rm\scriptscriptstyle [H]}_{\scriptscriptstyle{\rm{{ON}}} \,\scriptscriptstyle  i} (h)=0 	
		\\
		\bar{y}_{{\rm\scriptscriptstyle m}} \leq\textstyle\sum_{i=1}^{N_{\rm\scriptscriptstyle g}} q^{\rm\scriptscriptstyle [H]}_{{\scriptscriptstyle\rm g}\,\scriptscriptstyle i}(h)  \leq \bar{y}_{{\rm\scriptscriptstyle M}} 
		\\ \qquad 
		\text{otherwise }
	\end{array}\right.  \label{eq:y_glob}	
	\end{align}
\begin{align}
	\text{and,  }&\forall i=1,\dots,N_{\scriptscriptstyle\rm g} \nonumber\\
	& \text{MLD model of unit } i\nonumber\\
	& u_{{\rm\scriptscriptstyle m},i} x^{\rm\scriptscriptstyle [H]}_{\scriptscriptstyle{\rm{{ON}}} \,\scriptscriptstyle  i} (h) \leq   q^{\rm\scriptscriptstyle [H]}_{{\scriptscriptstyle\rm s}\,\scriptscriptstyle i}(h) \leq u_{{\rm\scriptscriptstyle M}, \scriptscriptstyle i} x^{\rm\scriptscriptstyle [H]}_{\scriptscriptstyle{\rm{{ON}}} \,\scriptscriptstyle  i} (h) \label{eq:u_i_cons_simpl}\\
	&y_{{\rm\scriptscriptstyle m},i} x^{\rm\scriptscriptstyle [H]}_{\scriptscriptstyle{\rm{{ON}}} \,\scriptscriptstyle  i} (h) + \gamma_{\scriptscriptstyle{\rm{{ST}}} \, \scriptscriptstyle  i} x^{\rm\scriptscriptstyle [H]}_{\scriptscriptstyle{\rm{{ST}}} \,\scriptscriptstyle  i}(h)\leq  y^{\rm\scriptscriptstyle [H]}_{\scriptscriptstyle i} (h) \nonumber\\
	& \leq y_{{\rm\scriptscriptstyle M},i} x^{\rm\scriptscriptstyle [H]}_{\scriptscriptstyle{\rm{{ON}}} \,\scriptscriptstyle  i} (h) + \gamma_{\scriptscriptstyle{\rm{{ST}}} \, \scriptscriptstyle  i} x^{\rm\scriptscriptstyle [H]}_{\scriptscriptstyle{\rm{{ST}}} \,\scriptscriptstyle  i}(h) \label{eq:y_i_cons_simpl}\\
	\text{ }&\forall h=0,\dots,N_{\scriptscriptstyle\rm H}\nonumber
	\end{align}
\end{subequations}
The decision variables are defined as a sequence of vectors along the optimization horizon, i.e., $\forall h=0,\dots,N_{\rm H}$: steam flow-rate, $\mathbf{q}^{\rm\scriptscriptstyle [H]}_{{\scriptscriptstyle\rm s}\,\scriptscriptstyle i}(h) =\linebreak[1] [{q}^{\rm\scriptscriptstyle [H]}_{{\scriptscriptstyle\rm s}\,\scriptscriptstyle i}(h),\linebreak[1] \dots,\linebreak[1] {q}^{\rm\scriptscriptstyle [H]}_{{\scriptscriptstyle\rm s}\,\scriptscriptstyle i}(h+N_{\rm\scriptscriptstyle H})]$ and the Boolean command for FSM transitions $\boldsymbol{\beta}^{\rm\scriptscriptstyle [H]}_{\scriptscriptstyle i}(h) =\linebreak[1] [\beta^{\rm\scriptscriptstyle [H]}_{\scriptscriptstyle i}(h),\linebreak[1] \dots,\linebreak[1] \beta^{\rm\scriptscriptstyle [H]}_{\scriptscriptstyle i}(h+N_{\rm\scriptscriptstyle H})]$ of each boiler, i.e. $\forall i={1,\dots, N_{\scriptscriptstyle\rm g}}$.\\
The cost function $J:\mathbb{R}^{\scriptscriptstyle n}\rightarrow\mathbb{R}$ is defined by summing the subsystems' stage costs $l_{\scriptscriptstyle i}(h)$ i.e., the operating cost related to the fuel consumption - based on natural gas price $\lambda_{\scriptscriptstyle{\rm{g}}}$ - fixed operating cost connected to the production mode $\lambda_{\scriptscriptstyle{\rm{ON}} \,\scriptscriptstyle  i}$ and the fixed startup costs $\lambda_{\scriptscriptstyle{\rm{ST}} \,\scriptscriptstyle  i}$.
The fixed costs are in general specific for each generator: they can include personnel, maintenance and degradation costs, that might increase for frequent start and stops.
\begin{equation}\label{eq:alpha_opt_descr}
\resizebox{.85\hsize}{!}{$\begin{array}{rl} l_{\scriptscriptstyle i}(h) =&\! \lambda_{\scriptscriptstyle{\rm{ON}} \,\scriptscriptstyle  i} (x^{\rm\scriptscriptstyle [H]}_{\scriptscriptstyle{\rm{{ON}}} \,\scriptscriptstyle  i} (h)) + \lambda_{\scriptscriptstyle{\rm{ST}}\, \scriptscriptstyle  i} (x^{\rm\scriptscriptstyle [H]}_{\scriptscriptstyle{\rm{{ST}}} \scriptscriptstyle  i} (h)) +  \\
\lambda_{{\scriptscriptstyle\rm{g}}}\frac{ T_{\scriptscriptstyle\rm H}}{\rho_{\scriptscriptstyle\rm g}}&\!\!  \left[ \left(g_i q^{\rm\scriptscriptstyle [H]}_{{\rm s}\, i}(h) \!+\! \gamma_{\scriptscriptstyle{\rm{ON}}\, \scriptscriptstyle  i}\right) x^{\rm\scriptscriptstyle [H]}_{\scriptscriptstyle{\rm{{ON}}}\, \scriptscriptstyle  i} (h)+\right.
\left.\gamma_{\scriptscriptstyle{\rm{{ST}}}\,  \scriptscriptstyle  i} x^{\rm\scriptscriptstyle [H]}_{\scriptscriptstyle{\rm{{ST}}}\, \scriptscriptstyle  i}(h)\right]\end{array}$}
\end{equation}%
Note that constraints \eqref{eq:u_glob}-\eqref{eq:y_glob} - enforced to guarantee \eqref{eq:uy_all} - are defined by logical conditions. A so-called ``Big-M'' reformulation can be adopted to transform these conditional constraints in a set of mixed-integer inequalities \cite{bemporad1999control}.\\
We denote by $\bar{\cdot}_{\rm\scriptscriptstyle m}$ ($\bar{\cdot}_{\rm\scriptscriptstyle M}$) the minimum (maximum) values of inputs and outputs, while $\bar{y}_{\rm\scriptscriptstyle ST} = \textstyle\sum_{i=1}^{N_{\rm\scriptscriptstyle g}} x^{\rm\scriptscriptstyle [H]}_{\scriptscriptstyle{\rm{{ST}}} \,\scriptscriptstyle  i} (h)\gamma_{\scriptscriptstyle{\rm{{ST}}}\,  \scriptscriptstyle  i}$.
At each step $h$, the optimizer computes the optimal trajectory of the sharing factors $\alpha^{\rm\scriptscriptstyle [H]}(j)$ for all $j=h,\dots,h+N_{\rm H}$. Based on the receding horizon principle, the  configuration $\alpha^{\rm\scriptscriptstyle [H]}(h)$, related to the first step, is broadcast to the network, while the rest of the trajectory is discarded (or, better said, kept as backup solution). At the subsequent step, $h+1$, the status of the GUs is retrieved, as well as an updated forecast of the future demand, moving forward the prediction horizon by one step.
This strategy permits to correct the demand forecast of remote steps as soon as they come closer, thus adjusting inaccurate estimations. A new profile $\alpha^{\rm\scriptscriptstyle [H]}(j)$, with $j=h+1,\dots,h+N_{\rm\scriptscriptstyle H}+1$, is computed by \eqref{eq:opt_simpl} and the solution  $\alpha^{\rm\scriptscriptstyle [H]}(h+1)$  sent to the GUs.
\begin{remark}
	The hard constraint \eqref{eq:qs_DemCon_simpl} can be tightened to equality, accelerating the solution convergence - if any feasible solution exists.
	Otherwise, if the program is infeasible, as the constraint  \eqref{eq:qs_DemCon_simpl} cannot be satisfied for certain demand profiles, it can be relaxed thanks to a slack variable $\varepsilon\geq 0$ with the modified objective function \eqref{eq:alpha_opt_simpl}, $\hat{l} = l + \lambda_{\varepsilon}\varepsilon^2$ with the constraint $\sum_{i=1}^{N_{ \rm g}} q^{\rm\scriptscriptstyle [H]}_{{\rm\scriptscriptstyle s}\,\scriptscriptstyle i}(h) \geq \bar{q}_{\rm\scriptscriptstyle s}^{\rm\scriptscriptstyle Dem}(h) -\varepsilon$. 
\end{remark}
\begin{remark}
		We solve~\eqref{eq:opt_simpl} in a centralized way, since the solution must be available with a frequency $f_{\rm\scriptscriptstyle H}=1/T_{\rm\scriptscriptstyle H}$. However, its computational complexity scales with the number of generation units, which can be very large in some applications. To overcome this, one may implement~\eqref{eq:opt_simpl} in a distributed fashion, 	as in \cite{falsone2018distributed}, partitioning the set of generators in clusters.
\end{remark}
\begin{remark}
	 The accuracy of demand forecast strongly impacts on the solution quality: since the reference is an additional decision variable, the feasibility is guaranteed. However, whenever the mismatch between demand forecast and its actual value is greater than a given threshold, the execution of the HL optimization can be triggered at an event-based "asynchronous" fashion to foster optimal tracking performances. 
\end{remark}%
A good demand forecast is indeed one of the main challenges for practical implementation of any scheme aiming to schedule the generation units. Small scale  generators, for medium-pressure steam, are usually operated in the industrial context where steam is considered a commodity resource. Therefore sometimes no demand forecast is available and neither considered for the generator management. Actually, accurate forecasts can be easily obtained by historical data and future production scheduling. Nowadays, companies that aim to implement energy efficiency strategies are increasing their awareness on energy utilization, through the analysis of historical data, and are pushed to implement procedures to correlate energy demand with production, giving the tools for deriving approximated evaluations of future steam demand to be used as input of the proposed management architecture. %
\subsection{Medium-level control} \label{sec:M-MPC}
The ML controller regulates the ensemble based on the operating modes and the sharing factors defined by the higher layer, driving the ensemble input $\bar{u}^{\rm\scriptscriptstyle [M]}(k)$ to the steady-state value, $\bar{u}^{\rm\scriptscriptstyle [H]}_{\rm ss}(h)=\sum_{i=1}^{N_{\rm g}} q^{\rm\scriptscriptstyle [H]}_{{\rm s}\, i}(h)$, computed by the HL optimizer. The medium-level MPC deals with an aggregate - small scale - model of the whole ensemble.
\subsubsection{Reference models and consistency requirements}
\label{subsec:Ref_model}
Medium level controller design requires, first of all, to devise an aggregate model of the ensemble. According to   \cite{Petzke2018}, a \emph{reference} model must be derived for each subsystem, defined as 
\begin{equation}
\resizebox{.85\hsize}{!}{$\hat{\mathcal{L}}_{\scriptscriptstyle i} : \left\lbrace  \begin{array}{rl}\hat{x}^{\rm\scriptscriptstyle [M]}_{i}(k+1) =& \hat{A} \hat{x}^{\rm\scriptscriptstyle [M]}_{i}(k) + \hat{B}_{\scriptscriptstyle i} u^{\rm\scriptscriptstyle [M]}_{i}(k)+\hat{w}^{\rm\scriptscriptstyle [M]}_{\scriptscriptstyle i}(k) \\ \hat{y}^{\rm\scriptscriptstyle [M]}_{i}(k) =&  \hat{C} \hat{x}^{\rm\scriptscriptstyle [M]}_{i}(k) + \hat{\gamma}_{\scriptscriptstyle i}\end{array} \right.$}
\label{eq:ref_model_i}
\end{equation}
where this  alternative model can be built on a possibly reduced state, defined as $\hat{x}^{\rm\scriptscriptstyle [M]}_{\scriptscriptstyle i}=\beta_{\scriptscriptstyle i} x^{\rm\scriptscriptstyle [M]}_{\scriptscriptstyle i}$, where $\beta_{\scriptscriptstyle i}\in \mathbb{R}^{\hat{n}\times n_{\scriptscriptstyle i}}$ is a suitable map, with $\hat{n}\leq n_{\scriptscriptstyle i}$. In addition, a term $\hat{w}^{\rm\scriptscriptstyle [M]}_{\scriptscriptstyle i}(k)$ is introduced to embed the error due to the mismatch between the reference model \eqref{eq:ref_model_i} and the identified system \eqref{eq:systems}.\\
By design, the state matrix $\hat{A}$ and the output matrix $\hat{C}$ can be generically defined: they just must be the same for all subsystems' reference models.  Conversely, the input matrix $\hat{B}_{\scriptscriptstyle i}$ must be accurately defined.
It is advantageous to select $\hat{A}$, $\hat{B}_{\scriptscriptstyle i}$ and $\hat{C}$ with the same canonical structure of ${A}_{\scriptscriptstyle i}$, ${B}_{\scriptscriptstyle i}$ and ${C}_{\scriptscriptstyle i}$, as defined in Section \ref{sec:sub:L_CL}.
Using this convenient choice, the state-reduction map $\beta_{\scriptscriptstyle i}\in \mathbb{R}^{\hat{n}\times n_{\scriptscriptstyle i}}$ is merely a selection matrix, whose rows are basis vectors of the  new canonical space. 
In this way the state of the reference models, is
$\hat{x}^{\rm\scriptscriptstyle [M]}(k) = [\delta y^{\rm\scriptscriptstyle [M]}({k}),\linebreak[1] \delta y^{\rm\scriptscriptstyle [M]}({k-1}),\linebreak[1]...,\linebreak[1] \delta y^{\rm\scriptscriptstyle [M]}({k-\hat{n}_{\rm f}}+1),\linebreak[1] {u}^{\rm\scriptscriptstyle [M]}({k-1}),\linebreak[1] ...,\linebreak[1] {u}^{\rm\scriptscriptstyle [M]}({k-\hat{n}_{\rm b}}+1)]^{\rm\scriptscriptstyle T}$.\\
The input matrix of the reference system must be defined in order to  satisfy the so-called \emph{gain consistency} conditions (see \cite{Petzke2018}): 
the reference model \eqref{eq:ref_model_i} and the  model \eqref{eq:systems} must guarantee to have the same static gain and a consistent output map. This is verified by imposing:
\begin{subequations}
\begin{align}
\hat{\gamma}_{\scriptscriptstyle i} =&\gamma_{\scriptscriptstyle{\rm{ON}} \,\scriptstyle i} \label{eq:gain_cons1}\\
\hat{b}_{\scriptscriptstyle i,1} =& \frac{\sum_{j=1}^{n_{\rm b}} b_{i,j} } {1+\sum_{j=1}^{n_{\rm f}} f_{\scriptscriptstyle i,j}} (1+ \sum_{j=1}^{\hat{n}_{\rm f}} \hat{f}_{\scriptscriptstyle j}) - \sum_{j=2}^{\hat{n}_{\rm b}} \hat{b}_{\scriptscriptstyle j} \label{eq:gain_cons2}
\end{align}
\label{eq:gain_cons_both}
\end{subequations}
 where $(b_{\scriptscriptstyle i,j},f_{\scriptscriptstyle i,j})$ and $(\hat{b}_{\scriptscriptstyle i,j}, \hat{f}_{\scriptscriptstyle i,j})$ are the parameters of the $i$-th models \eqref{eq:systems} and \eqref{eq:ref_model_i}, respectively.
\subsubsection{Disturbance $\hat{w}^{\rm\scriptscriptstyle [M]}_{\scriptscriptstyle i}(k)$}
As discussed, the term $\hat{w}^{\rm\scriptscriptstyle [M]}_{\scriptscriptstyle i}(k)$ embeds the error due to the mismatch between the reference model \eqref{eq:ref_model_i} and the original one \eqref{eq:systems}  induced by the selection of the same state matrices for the reference models. To apply robust MPC for ensemble control, we need to ensure that $\hat{w}^{\rm\scriptscriptstyle [M]}_{\scriptscriptstyle i}(k)$ is bounded.
In \cite{Petzke2018}, it is shown that the set where  $\hat{w}^{\rm\scriptscriptstyle [M]}_{\scriptscriptstyle i}(k)$ lies (i.e., $\mathcal{W}_{\scriptscriptstyle i}$) can be made small by properly restricting the set of $\Delta{u}^{\rm\scriptscriptstyle [M]}_{\scriptscriptstyle i}(k) = {u}^{\rm\scriptscriptstyle [M]}_{\scriptscriptstyle i}(k) - {u}^{\rm\scriptscriptstyle [M]}_{\scriptscriptstyle i}(k-1)$, i.e., $\Delta\bar{\mathcal{U}}_{\scriptscriptstyle i}$. However, the definition of $\mathcal{W}_{\scriptscriptstyle i}$ used in \cite{Petzke2018} requires the definition of a suitable invariant set, used to define the low-level MPC controller, which is here absent.\\
In any case the fact that $\mathcal{W}_{\scriptscriptstyle i}$ depends upon $\Delta\bar{\mathcal{U}}_{\scriptscriptstyle i}$ remains valid also in this framework, i.e., when the low-level controller is unconstrained. This is supported by the fact that
$\hat{w}_i(k)=\beta_ix_i(k+1)\linebreak[1]-\hat{x}_i(k+1)=\linebreak[1] \beta_i [\delta y_i({k+1}),\linebreak[1] \delta y_i({k}),\linebreak[1]...,\linebreak[1] \delta y_i({k-n_{\rm f}}+2),\linebreak[1] {u}_i({k}),\linebreak[1] ...,\linebreak[1] {u}_i({k-n_{\rm b}}+2)]^T\linebreak[1]-[\delta y^o_i({k+1}),\linebreak[1] \delta y^o_i({k}),\linebreak[1]...,\linebreak[1] \delta y^o_i({k-\hat{n}_{\rm f}}+2),\linebreak[1] {u}_i({k}),\linebreak[1] ...,\linebreak[1] {u}_i({k-\hat{n}_{\rm b}}+2)]^T\linebreak[1]=
[\delta y_i({k+1})-\delta y_i^o({k+1}),\linebreak[1] \delta y_i({k})-\delta y_i^o({k}),\linebreak[1]...,\linebreak[1] \delta y_i({k-\hat{n}_{\rm f}}+2)-\delta y_i^o({k-\hat{n}_{\rm f}}+2),\linebreak[1] 0,\linebreak[1] ...,\linebreak[1]0]^T$, where $\delta y_i^o(k)$ is defined as the output of the "unperturbed" reference system
\begin{equation}
\hat{\mathcal{S}}_i^o : \left\lbrace  \begin{array}{rl}\hat{x}^o_{i}(k+1) =& \hat{A} \hat{x}^o_{i}(k) + \hat{B}_i u_{i}(k) \\ \delta\hat{y}^o_{i}(k) =&  \hat{C} \hat{x}^o_{i}(k)\end{array} \right.
\label{eq:ref_model_i_unp}\end{equation}
In view of this, each non-zero component of vector $\hat{w}_i(k)$ is a lagged version of $e_y(k+1)=\delta y({k+1})-\delta y^o({k+1})$. It is possible to show that, thanks to the gain consistency condition, there exists a transfer function $\Delta \mathcal{G}_i(z^{-1})$ such that\footnote{To retrieve \eqref{eq:du-e} we can write, from \eqref{eq:systems} and~\eqref{eq:ref_model_i_unp}, that
	$$e_i(k+1)= G_i(z^{-1}) u_i(k) -\hat{G}_i(z^{-1}) u_i(k)$$
	where $G_i$ and $\hat{G}_i$ are the transfer functions of systems \eqref{eq:systems},~\eqref{eq:ref_model_i_unp}, respectively, while $z^{-1}$ is the discrete time backward shifting operator. According to the gain consistency property, $\hat{G}_i(1)=G_i(1)$. Therefore
	$$e_i(k+1)=\left( (G_i(z^{-1}) -G_i(1) )-(\hat{G}_i(z^{-1})-\hat{G}_i(1)) \right) u_i(k)$$
	Considering the canonical structure of the model \eqref{eq:systems},
	the term $ {G}_i(z^{-1})-{G}_i(1)$ has the following form:
	$${G}_i(z^{-1})-{G}_i(1) = \frac{\sum_{j=1}^{n_{\rm b}}(b_{i,j}(z^{-j}-1))}{\sum_{j=1}^{n_{\rm f}}(f_{i,j}(z^{-j}-1))}$$
	By the rational root theorem, every binomial $(z^{-j}-1)$ can be factorized by $(z^{-1}-1)$, therefore we can write
	$${G}_i(z^{-1})-{G}_i(1) =(z^{-1}-1)\mathcal{G}_i(z^{-1})$$ and, similarly, 
	$$\hat{G}_i(z^{-1})-\hat{G}_i(1) =(z^{-1}-1)\hat{\mathcal{G}}_i(z^{-1})$$
	Therefore,
	\begin{align}e_i(k+1) &=\left[\mathcal{G}_i(z^{-1})- \hat{\mathcal{G}}_i(z^{-1}) \right](z^{-1}-1)u_i(k)\nonumber\\ 
	&= -\Delta \mathcal{G}_i(z^{-1}) (u_i(k) - u_i(k-1))\nonumber \end{align} }
\begin{equation}e_i(k+1)= -\Delta \mathcal{G}_i(z^{-1}) \Delta u_i(k)\label{eq:du-e}\end{equation}
Following \cite{kolmanovsky1998theory}, the set $\mathcal{W}_{\scriptscriptstyle i}$ can be explicitly computed based on \eqref{eq:du-e}. However, in this work, to quantify set $\mathcal{W}_{\scriptscriptstyle i}$, due to its convexity, we have taken the convex hull of the points - given simulating with a signal $\Delta u_i(k)$ sampled from  $\Delta\bar{\mathcal{U}}_i$ - to approximate the set. This solution has permitted to apply the robust MPC approach defined in this section with no constraint violation on the real variables.
\subsubsection{Ensemble model}
\label{subsec:Ensemle_model}
To define the ensemble dynamics, the reference models must be opportunely combined. 
The state of the ensemble dynamical model $\bar{\mathcal{L}}$ is composed of the states of the active generators,  i.e., with $x^{\rm\scriptscriptstyle [H]}_{\scriptscriptstyle{\rm{{ON}}} \,\scriptstyle i}=1$. When a boiler is switched off, its contribution to ensemble steam production is immediately removed: in practice, during the transient, its steam is diverted from the ensemble output. Similarly, during start-up, the produced steam is not conveyed to the ensemble output, due to low steam quality - with high percentage of transported condensate.
Accordingly, we define the ensemble state as $\bar{x}^{\rm\scriptscriptstyle [M]} = \sum_{\scriptscriptstyle i}^{N_{\rm g}} x^{\rm\scriptscriptstyle [H]}_{\scriptscriptstyle{\rm{{ON}}} \,\scriptstyle i}\hat{x}^{\rm\scriptscriptstyle [M]}_{\scriptscriptstyle i}$, its input as  $\bar{u}^{\rm\scriptscriptstyle [M]}$, and its output as $\bar{y}^{\rm\scriptscriptstyle [M]} = \sum_{\scriptscriptstyle i}^{N_{\rm g}} x^{\rm\scriptscriptstyle [H]}_{\scriptscriptstyle{\rm{{ON}}} \,\scriptstyle i}  \hat{y}^{\rm\scriptscriptstyle [M]}_{\scriptscriptstyle i}$.\\ Considering the reference models \eqref{eq:ref_model_i}, we can write
\begin{equation}
\resizebox{.85\hsize}{!}{$\bar{\mathcal{L}} : \left\lbrace  \begin{array}{r l}{\bar{x}^{\rm\scriptscriptstyle [M]}}(k+1) =& \hat{A}\bar{x}^{\rm\scriptscriptstyle [M]}(k)+ \bar{B} {\bar{u}}^{\rm\scriptscriptstyle [M]}(k )+\bar{w}^{\rm\scriptscriptstyle [M]}(k)\\ \bar{y}^{\rm\scriptscriptstyle [M]}(k) = & \hat{C} \bar{x}^{\rm\scriptscriptstyle [M]}(k) + \bar{\gamma} \end{array} \right.$}
\label{eq:ensemble_model}
\end{equation}
where $\bar{B} = \sum_{\scriptscriptstyle i}^{N_{\rm g}} \alpha^{\rm\scriptscriptstyle [H]}_{\scriptscriptstyle i} \hat{B}_{\scriptscriptstyle i}$, $\bar{\gamma} = \sum_{\scriptscriptstyle i}^{N_{\rm g}} x^{\rm\scriptscriptstyle [H]}_{\scriptscriptstyle{\rm{{ON}}} \,\scriptstyle i} \hat{\gamma}_{\scriptscriptstyle i}$, and $\bar{w}^{\rm\scriptscriptstyle [M]}=\sum_{\scriptscriptstyle i}^{N_{\rm g}}x^{\rm\scriptscriptstyle [H]}_{\scriptscriptstyle{\rm{{ON}}} \,\scriptstyle i} \hat{w}^{\rm\scriptscriptstyle [M]}_{\scriptscriptstyle i}$.
We also define the static gain of the ensemble as $\bar{g} = \sum_{\scriptscriptstyle i}^{N_{\rm g}} \alpha^{\rm\scriptscriptstyle [H]}_{\scriptscriptstyle i} g_{\scriptscriptstyle i}$.
\begin{remark}
The gain consistency conditions \eqref{eq:gain_cons_both} are necessary to guarantee that the ensemble gain correctly reflects the overall gains of the subsystems, given the specified load partition.	
\end{remark}
The set containing the reference deviation $\bar{w}^{\rm\scriptscriptstyle [M]}(k)$ is defined as $\bar{\mathcal{W}}$. It can be computed as  discussed in \cite{Petzke2018}. More specifically,  we can enforce - as discussed in Section \ref{subsec:MPC} - $\Delta u^{\rm\scriptscriptstyle [M]}_{\scriptscriptstyle i}(k)\in\Delta \bar{\mathcal{U}}$, for all $i=1,\dots,N_{\rm\scriptscriptstyle g}$ and for all values of $\alpha^{\rm\scriptscriptstyle [H]}_{\scriptscriptstyle i}$, where $\Delta \bar{\mathcal{U}}=[-\Delta \bar{u},\Delta \bar{u}]$ for a given threshold $\Delta\bar{u}$. As discussed, this is done to guarantee that  $\bar{w}^{\rm\scriptscriptstyle [M]}_{\scriptscriptstyle i}(k)\in\mathcal{W}_{\scriptscriptstyle i}$, and also that $\bar{w}^{\rm\scriptscriptstyle [M]}(k)\in\bar{\mathcal{W}}=\bigoplus_{i=1}^{N_{\rm\scriptscriptstyle g}}\mathcal{W}_{\scriptscriptstyle i}$ in all possible system configurations. 
\subsubsection{Medium-level controller design} \label{subsec:MPC}
The ML MPC objective is to track the global fuel flow-rate target $r=\bar{q}_{\scriptscriptstyle\rm g}^{\rm\scriptscriptstyle Dem}$, that depends on the HL solution. %
At any time instant $k$, the HL share and mode signals, $(\alpha^{\rm\scriptscriptstyle [H]}_{\scriptscriptstyle i}, x^{\rm\scriptscriptstyle [H]}_{{\scriptscriptstyle m}\,i})$, are re-sampled with sampling time $T_{\rm\scriptscriptstyle M}$, as $\alpha^{\rm\scriptscriptstyle [M]}_{\scriptscriptstyle i}(k)=\alpha^{\rm\scriptscriptstyle [H]}_{\scriptscriptstyle i}(\lfloor k/\mu\rfloor)$,  
and are assumed to remain constant, e.g., $\alpha^{\rm\scriptscriptstyle [M]}_{\scriptscriptstyle i}(k+l)=\alpha^{\rm\scriptscriptstyle [M]}_{\scriptscriptstyle i}(k)$
for the whole control horizon, i.e.,  $\forall l=1,\dots,N_{\scriptscriptstyle\rm M}$. This implies that the ensemble model $\bar{\mathcal{L}}$,~\eqref{eq:ensemble_model}, is invariant during the optimization horizon, $N_{\scriptscriptstyle\rm M}$.\\
To cope with disturbance $\bar{w}^{\rm\scriptscriptstyle [M]}(k)$ in the ensemble model $\bar{\mathcal{L}}$, the ML must be designed according to a robust tube-based implementation. The system is augmented and written in velocity form, as in~\cite{Betti2013}
\begin{equation}
\xi^{\rm\scriptscriptstyle [M]}(k+1) = \mathcal{A} \xi^{\rm\scriptscriptstyle [M]}(k) + \mathcal{B} \Delta \bar{u}^{\rm\scriptscriptstyle [M]}(k) + \mathcal{H} \Delta \bar{w}^{\rm\scriptscriptstyle [M]}(k) 
\label{eq:aug_model_act}
\end{equation}	
with state vector $\xi^{\rm\scriptscriptstyle [M]}(k)=[\Delta \bar{x}^{\rm\scriptscriptstyle [M]}(k), \varepsilon^{\rm\scriptscriptstyle [M]}(k)]$, input $\Delta \bar{u}^{\rm\scriptscriptstyle [M]}(k)$, and disturbance $\Delta \bar{w}^{\rm\scriptscriptstyle [M]}(k)$. Matrices $\mathcal{A, B, H}$ can be trivially derived from \eqref{eq:ensemble_model}.\\
The added state is $\varepsilon^{\rm\scriptscriptstyle [M]}(k)= \bar{y}^{\rm\scriptscriptstyle [M]}(k) - \hat{r}$, where the reference output $\hat{r}$  is set as a decision variable of the OCP, as in \cite{Limon2008}, to ensure recursive feasibility and offset-free tracking capabilities in presence of continuous variations of the target values (which can be possibly infeasible): in a few words, $\hat{r}$ is the closest feasible set-point to $r$, at least in stationary conditions.\\
A nominal (undisturbed) model, used to formulated the OCP, can be associated to \eqref{eq:aug_model_act}:
\begin{equation}
	\tilde{\xi}^{\rm\scriptscriptstyle [M]}(k+1) = \mathcal{A} \tilde{\xi}^{\rm\scriptscriptstyle [M]}(k) + \mathcal{B} \Delta \tilde{u}^{\rm\scriptscriptstyle [M]}(k) 
	\label{eq:aug_model_nom}
\end{equation}	
whose variables are denoted by $\tilde{\cdot}$.\\
To guarantee the feasibility in the disturbed case, the constraints for the OCP with nominal model must be opportunely tightened. 
The tube-based approach requires the computation of a Robust Positively Invariant (RPI) set ${\mathcal{Z}}$ - computed based on \cite{rakovic2005invariant} - where $\xi^{\rm\scriptscriptstyle [M]}(k)-\tilde{\xi}^{\rm\scriptscriptstyle [M]}(k)$ is guaranteed to lie if the following control law is applied to the real system, 
\begin{equation}{\delta{\bar{u}}^{\rm\scriptscriptstyle [M]}}(k) = {\delta\tilde{u}^{\rm\scriptscriptstyle [M]}}(k) + \mathcal{K} (\xi^{\rm\scriptscriptstyle [M]}(k)-\tilde{\xi}^{\rm\scriptscriptstyle [M]}(k)) \label{eq:u_corr}
\end{equation}
where $\mathcal{K}$ a gain matrix that makes the matrix $\mathcal{A}+\mathcal{B}\mathcal{K}$ Schur stable. Namely, the real system is kept close to the nominal state, i.e.,
$$\xi^{\rm\scriptscriptstyle [M]}(k+j) \in \tilde{\xi}^{\rm\scriptscriptstyle [M]}(k) \oplus {\mathcal{Z}} \qquad \forall j \geq 1 $$ 
So, the robust MPC problem is formulated on the nominal system \eqref{eq:aug_model_nom}, leading to a quadratic program (QP), where the optimization variables are the future nominal input trajectory, $\delta\mathbf{\tilde{u}}(k) = [\delta{\tilde{u}}^{\rm\scriptscriptstyle [M]}(k):\delta{\tilde{u}}^{\rm\scriptscriptstyle [M]}(k+N_{\rm \scriptscriptstyle M}-1)]$, the initial condition of the nominal system, $\tilde{\xi}^{\rm\scriptscriptstyle [M]}(k) = (\delta\tilde{x}^{\rm\scriptscriptstyle [M]}(k),\tilde{y}^{\rm\scriptscriptstyle [M]}(k)-\hat{r})$, and the output reference point, $\hat{r}$.
\begin{subequations}
	\begin{equation} \label{eq:cost_MPC}
	\min_{\substack{\tilde{\xi}^{\rm\scriptscriptstyle [M]}(k), \hat{r},\\ \delta\mathbf{\tilde{u}}(k)}} \,
	\|\hat{r} - r \|_{\scriptscriptstyle T}^{\scriptscriptstyle 2} + \sum_{j\in\mathcal{J}} \left\lbrace \| \tilde{\xi}^{\rm\scriptscriptstyle [M]}(j)\|_{\scriptscriptstyle Q}^{\scriptscriptstyle 2} + \| \delta{\tilde{u}}^{\rm\scriptscriptstyle [M]}(j)\|_{\scriptscriptstyle R}^{\scriptscriptstyle 2}\right\rbrace
\end{equation}
\begin{align}
\text{s.t.} \quad 
& {\xi}^{\rm\scriptscriptstyle [M]}(k)-\tilde{\xi}^{\rm\scriptscriptstyle [M]}(k)
\in {\mathcal{Z}}& \label{eq:const_init_MPC}\\
&\tilde{\xi}^{\rm\scriptscriptstyle [M]}(j+1) = \mathcal{A}\tilde{\xi}^{\rm\scriptscriptstyle [M]}(j) + \mathcal{B}{\delta\tilde{u}^{\rm\scriptscriptstyle [M]}}(j) & \label{eq:const_sysdyn_MPC}\\
&\tilde{u}^{\rm\scriptscriptstyle [M]}(j)  \in \tilde{\mathcal{U}}&\label{eq:constr1_MPC}\\
& \alpha^{\rm\scriptscriptstyle [H]}_{\scriptscriptstyle i}(h)  \tilde{u}^{\rm\scriptscriptstyle [M]}(j) \in \tilde{\mathcal{U}}_{\scriptscriptstyle i}&
\label{eq:constr3_MPC}\\
& x_{\scriptscriptstyle{\rm{\scriptscriptstyle {ON}}} \,\scriptscriptstyle  i}^{\rm\scriptscriptstyle [H]} (h) \left[ g_{\scriptscriptstyle i} \alpha_{\scriptscriptstyle i}(h) \tilde{u}^{\rm\scriptscriptstyle [M]}(j) +\hat{\gamma}_{\scriptscriptstyle{\rm{\scriptscriptstyle ON}} \,\scriptscriptstyle  i}  \right] \in \tilde{\mathcal{Y}}_i& \label{eq:constr4_MPC}\\
&\alpha^{\rm\scriptscriptstyle [M]}_{\scriptscriptstyle i}(j) \tilde{u}^{\rm\scriptscriptstyle [M]}(j)- \alpha^{\rm\scriptscriptstyle [M]}_{\scriptscriptstyle i}(j-1)\tilde{u}^{\rm\scriptscriptstyle [M]}(j-1)\in \Delta \tilde{\mathcal{U}} \label{eq:constr5_MPC}\\
&\quad \begin{array}{l}\quad\forall j\in\mathcal{J} \\\quad \forall i=1, \dots, N_{\rm \scriptscriptstyle g}\end{array}\nonumber\\
&  \tilde{\xi}^{\rm\scriptscriptstyle [M]}(k+N_{\rm \scriptscriptstyle M})  = 0 \label{eq:const_xi_MPC2}\\
& \tilde{x}^{\rm\scriptscriptstyle [M]}(k+N_{\rm \scriptscriptstyle M}) =  \tilde{x}_{\scriptscriptstyle \rm ss} \label{eq:const_xN_MPC2}\\
& \tilde{u}^{\rm\scriptscriptstyle [M]}(k+N_{\rm \scriptscriptstyle M}-1) =  \tilde{u}_{\scriptscriptstyle \rm ss} \label{eq:const_uN_MPC2}
\end{align}%
\label{eq:MPC_Rob1}
\end{subequations}%
\noindent where $\mathcal{J}=\{k, \dots, k+N_{\rm \scriptscriptstyle M}-1\}$.
Moreover, $\tilde{x}_{\rm ss},$ $\tilde{u}_{\rm ss}$ are given by 
$$ \left[\arraycolsep=.5pt\def\arraystretch{1.1} \begin{array}{c} \tilde{x}_{\scriptscriptstyle \rm ss}\\ \tilde{u}_{\scriptscriptstyle \rm ss}\end{array}\right] =
\left[ \begin{array}{cc} I_{\scriptscriptstyle n}-\hat{A} & -\bar{B}\\ \hat{C} & 0_{\scriptscriptstyle m}\end{array}\right]^{\scriptscriptstyle -1}
\left[\arraycolsep=.5pt\def\arraystretch{1.1} \begin{array}{c} 0_{\scriptscriptstyle n\times p}\\ I_{\scriptscriptstyle p}\end{array}\right] (\hat{r}-\bar{\gamma})$$ 
The constraints \eqref{eq:const_xN_MPC2}-\eqref{eq:const_uN_MPC2} requires the calculation of $\tilde{x}^{\rm\scriptscriptstyle [M]}(k-1)$, $\tilde{u}^{\rm\scriptscriptstyle [M]}(k-1)$ that can be evaluated based on 
$$\left[\arraycolsep=.5pt\def\arraystretch{1.1} \begin{array}{c} \tilde{x}^{\rm\scriptscriptstyle [M]}(k-1)\\ \tilde{u}^{\rm\scriptscriptstyle [M]}(k-1)\end{array}\right] =
\left[\begin{array}{cc} \hat{A}- I_n & \bar{B} \\\hat{C}\hat{A} & \hat{C}\bar{B}\end{array}\right]^{\scriptscriptstyle -1}\left[\arraycolsep=.5pt\def\arraystretch{1.1} \begin{array}{c} \Delta\tilde{x}^{\rm\scriptscriptstyle [M]}(k)\\ \tilde{y}^{\rm\scriptscriptstyle [M]}(k)\end{array}\right]$$
Differently from \cite{Betti2013}, the terminal constraint is a \emph{steady-state} condition for \eqref{eq:aug_model_nom} in the last step of the prediction horizon. The computation of a terminal steady-state condition guarantees that the MPC problem is practically recursively feasible, with auxiliary control law ${\Delta{\tilde{u}}^{\rm\scriptscriptstyle [M]}}(k) =  0$.
This formulation avoids the computation of the Maximal Output Admissible Set (MOAS) required in \cite{Betti2013}. It is worth noting that - similarly to the computation of the RPI \cite{rakovic2005invariant} - the calculation of the MOAS \cite{gilbert1991linear} is an iterative time-consuming procedure. Any variation of the configuration requires the re-computation online of both the RPI and the MOAS. At least the latter is avoided by forcing the system to reach a steady-state condition at the end of the prediction window; on the other hand, it might affect the promptness of the controller, reducing the optimal control action, since ${\Delta{\tilde{u}}^{*[T]}}(k) \rightarrow  0$ as $k \rightarrow N_{\rm \scriptscriptstyle M}$. This can be mitigated by selecting a longer prediction window.\\
The initial condition of the nominal state is enforced by \eqref{eq:const_init_MPC}.
For all the time steps $j\in\mathcal{J}$, the ML is committed to impose the constraints \eqref{eq:constraints_all} through \eqref{eq:constr1_MPC}-\eqref{eq:constr4_MPC}. Moreover, as discussed in Section~\ref{subsec:Ref_model}, in order to keep the disturbance term $\bar{w}^{\rm\scriptscriptstyle [M]}(k)$ bounded, we need to ensure that for each generator the input variation is limited thanks to (the tightened) constraint \eqref{eq:constr5_MPC}.\\ 
Also constraints  \eqref{eq:constr1_MPC}-\eqref{eq:constr5_MPC} are imposed on the nominal system variables: this requires a proper tightening \cite{Betti2013} of the original sets $\bar{\mathcal{U}}$,  $\bar{\mathcal{U}}_i$,$\Delta\bar{\mathcal{U}}_{\scriptscriptstyle i}$ allowing us to define $\tilde{\mathcal{U}}$, $\tilde{\mathcal{U}}_i$, and $\Delta\tilde{\mathcal{U}}_{\scriptscriptstyle i}$.\\
Note also that, while in \cite{Petzke2018} constraints on local outputs are not considered, in our application scenario they play a key role. In fact they represent limitations in the gas available to each burner. To enforce $y_{\scriptscriptstyle i}^{\rm\scriptscriptstyle [M]}\in\mathcal{Y}_{\scriptscriptstyle i}$, we use its simplified ``quasi steady-state'' version \eqref{eq:constr4_MPC}. 
Set ${\tilde{\mathcal{Y}}}_{\scriptscriptstyle i}$ is computed by suitably tightening set ${\mathcal{Y}}_{\scriptscriptstyle i}$. 
\subsubsection{Transitions among configurations} \label{subsec:Trans_MPC}
When configuration transitions occur, i.e., when the high hierarchical level returns a new optimal value of sharing factor $\alpha_{\scriptscriptstyle i}^{*\rm\scriptscriptstyle [M]}(k)\neq \alpha^{*\rm\scriptscriptstyle [M]}_{\scriptscriptstyle i}(k-1)$ at least for some subsystems, infeasibility issues may occur due to two reasons: (i) the ensemble model is varying with respect to the one used at the previous time step,  since $\bar{B}=\bar{B}\left( \alpha^{*\rm\scriptscriptstyle [M]}(k)\right)$; (ii) it is not guaranteed that constraints \eqref{eq:constr1_MPC} and \eqref{eq:constr5_MPC} can be enforced in a recursive manner. 
The procedure adopted when configuration changes occur is the following one:
\begin{itemize}
	\item Apply  $\alpha^{\rm\scriptscriptstyle [M]}_{\scriptscriptstyle i}(k)=\alpha_{\scriptscriptstyle i}^{*\rm\scriptscriptstyle [M]}(k)$,  $\forall i=1,\dots,N_{\rm\scriptscriptstyle g}$, and solve the corresponding MPC optimization problem. If it is feasible, then the configuration change is accepted.
	\item If the optimization problem formulated at the previous time step does not result feasible, then reformulate the MPC optimization problem using the actual sharing factors ${\alpha}^{\rm\scriptscriptstyle [M]}(k)$ (under the assumption to keep them constant during the whole control horizon) as further - temporary - optimization variables and adding the term $\sum_{i=1}^{N_{\rm g}}\|{\alpha}^{\rm\scriptscriptstyle [M]}_{\scriptscriptstyle i}(k)-\alpha^{*\rm\scriptscriptstyle [M]}_{\scriptscriptstyle i}(k)\|^2$ to the cost function, in order to steer $\alpha_{\scriptscriptstyle i}^{\rm\scriptscriptstyle [M]}(k)$ to the values $\alpha^{*\rm\scriptscriptstyle [M]}_{\scriptscriptstyle i}(k)$, selected as the optimal ones by the high-level optimizer. 
\end{itemize}
\begin{remark}
The introduction of the sharing factors as additional decision variables transforms the program \eqref{eq:MPC_Rob1} from QP to a nonlinear one. In fact, the dependence of the model on  $\alpha_{\scriptscriptstyle i}^{\rm\scriptscriptstyle [M]}$ implies that a number of elements of problem \eqref{eq:MPC_Rob1} are dependent upon $\alpha_{\scriptscriptstyle i}^{\rm\scriptscriptstyle [M]}$ in a non-trivial way, e.g., the gain $\mathcal{K}$, the RPI set ${\mathcal{Z}}$ (to be used in the constraint \eqref{eq:const_init_MPC}), and the set tightening.\\ 
We can here address this issue by reformulating \eqref{eq:MPC_Rob1} in a slightly different, but consistent, way, to be applied exclusively during the transitions. First of all, to avoid the use of ${\mathcal{Z}}$, we replace \eqref{eq:const_init_MPC} with the equality $\tilde{\xi}^{\rm\scriptscriptstyle [M]}(k)={\xi}^{\rm\scriptscriptstyle [M]}(k)$. Also, due to Assumption \ref{ass:ass1}, $\hat{A}$ is Schur stable. Thus, we can adopt, during the transition, an auxiliary law with $\mathcal{K}=0$. So, the input applied to the model ensemble is not corrected by \eqref{eq:u_corr}. A final remark is in order: to support transitions, the tightening operations to be performed on sets $\bar{\mathcal{U}}$, $\Delta\bar{\mathcal{U}}$, $\bar{\mathcal{U}}_i$, and $\bar{\mathcal{Y}}_i$ should be sufficiently general to be compatible with all ensemble models of interest to avoid possible feasibility losses.
\end{remark}

\section{Simulations}
\label{sec:Simu}
The hierarchical control scheme is tested in simulation, considering a use-case with $N_{\rm\scriptscriptstyle g}=5$ steam generators that operate at a pressure of $57$ bar and cooperate to serve a common load. The boilers that form the ensemble are slightly different among each other, since they are characterized by dissimilar dimensions and efficiencies. Also, they are limited to work in different operating ranges, i.e. minimum/maximum generated steam. Their parameters are reported in Table \ref{tb:Boiler_par_1}.\\
\begin{table}[htb]
	\centering	
	\caption{Boiler parameters.}
	\scriptsize
	\begin{tabular}{r||c| c| c| c| c}		
		
		\textbf{Boiler n} & 1&2&3&4&5\\
		\hline
		$V_{\rm\scriptscriptstyle T}$ $[m^3]$& 1.21& 1.15& 1.28& 1.14& 1.32\\
		$M_{\rm\scriptscriptstyle T}$ $ [t]$& 5.49& 5.22& 5.83& 5.06& 5.99 \\
		$\eta$ $[-]$& 0.90& 0.92& 0.89& 0.95& 0.99\\
		$q_{\rm\scriptscriptstyle s}^{\rm\scriptscriptstyle Min}$   $[kg/s]$ & 0.1 & 0.09& 0.09& 0.09& 0.1\\
		$q_{\rm\scriptscriptstyle s}^{\rm\scriptscriptstyle Max}$ $[kg/s]$ & 1.26& 1.16& 1.13& 1.20& 1.25\\
		$q_{\rm\scriptscriptstyle g}^{\rm\scriptscriptstyle Min}$  $[kg/s]$ & 0.125  &  0.127  &  0.129  &  0.126  &  0.123\\
		$q_{\rm\scriptscriptstyle g}^{\rm\scriptscriptstyle Max}$  $[kg/s]$ & 0.859  &  0.844  &  0.846  &  0.841  &  0.839\\
		$\lambda_{\rm\scriptscriptstyle g}$ $[\text{\euro}/m^{3}]$ & 0.22  & 0.22 &  0.22 &   0.22 &   0.22\\
		$\lambda_{\scriptscriptstyle\rm{ON}}$ $[\text{\euro}/T_{\rm H}]$ & 40  & 30 &  22 &   55 &   45\\
		$\lambda_{\scriptscriptstyle\rm{ST}}$ $[\text{\euro}/T_{\rm H}]$ & 100  & 130 &  120 &   70 &   80\\
	\hline
	\end{tabular}
\label{tb:Boiler_par_1}
	\end{table}
The natural gas price $\lambda_{\scriptscriptstyle{\rm{g}}}$ is assumed fixed and constant for all the generators, while the fixed operating cost in ON and startup modes $\lambda_{\scriptscriptstyle{\rm{ON}} \,\scriptstyle i}$ and  $\lambda_{\scriptscriptstyle{\rm{ST}} \,\scriptstyle i}$, respectively,  are different for each generator. Gas density is $\rho_{\scriptscriptstyle\rm g}=0.71$ $[kg/m^{\scriptscriptstyle 3}]$ and tube specific heat $c_{\scriptscriptstyle\rm p}=0.5$ $[kJ/(kg K)]$.\\
The system is characterized by the following global constraints 
$\bar{\mathcal{Y}} = [0.1227, 4.220]$ $[kg/s]$ and $\bar{\mathcal{U}} = [0.089, 6.0]$ $[kg/s]$, determined  by constraints of the distribution network.\\
The sampling times of the multi-layer architecture are reported in Table \ref{tb:Boiler_par_2}.\\
\begin{table}[htb]
	\centering	
	\caption{Multi-layer sampling times.}
	\scriptsize
	\begin{tabular}{r||c|c|c}	
			
		\textbf{Sampling time} & $\tau$ & $T_{\scriptscriptstyle\rm M}$ & $T_{\scriptscriptstyle\rm H}$\\
		\hline
		& $10$ s & $30$ s & $10$ min \\
		\hline
	\end{tabular}
	\label{tb:Boiler_par_2}
\end{table}%
The low-level controllers have been implemented in discrete-time with a fast sampling time $\tau=10$ s; their parameters are tuned to stabilize the system with a settling time of $120$ s.\\ All systems are assumed to have the same compensator  $\mathbf{C}$, with $K_{\scriptscriptstyle\rm{P}}=0.30$ and $K_{\scriptscriptstyle\rm{I}}=0.10$ and regulator $\mathbf{R}$, with $K_{\scriptscriptstyle\rm{P}}=0.87$ and $K_{\scriptscriptstyle\rm{I}}=3.5\cdot{10}^{\scriptscriptstyle -4}$.\\ 
The discrete-time linear model \eqref{eq:poly_id} is identified on a data-set generated by simulating the closed-loop nonlinear model $\mathcal{S}_{i}^{\scriptscriptstyle \rm{CL}}$, with sampling time $T_{\scriptscriptstyle\rm M}=30$ s.
For each boiler, the identified models, $\mathcal{L}_{i}^{\scriptscriptstyle \rm{CL}}$, are characterized by $n_{\scriptscriptstyle\rm f}= 3$, $n_{\scriptscriptstyle\rm b}= 2$, and $n_{\scriptscriptstyle\rm k}= 1$. So that systems $\mathcal{L}_{i}^{\scriptscriptstyle \rm{CL}}$ have the same order $n$.  
The high-level optimization is executed in receding horizon with a slow sampling time $T_{\rm\scriptscriptstyle H} = 10$ min.\\
The optimization \eqref{eq:opt_simpl} considers a prediction horizon of $N_{\scriptscriptstyle\rm H} = 10$, which is long enough to consider the high-level dynamics of the sub-systems - by considering their start-up dynamics - and forthcoming fluctuation of the users global demand, $\bar{q}_{\scriptscriptstyle\rm s}^{\scriptscriptstyle\rm Dem}(h)$ for $h=0,\dots,N_{\scriptscriptstyle\rm H}$.\\
The latter is given as a piece-wise constant forecast of users' demand, which can be opportunely updated at any iteration of the rolling window of the high-level optimization.\\ 
Regarding the high-level dynamic models of the steam generators, each unit is characterized by an hybrid automaton, as presented in Section \ref{sec:Model-sub:HL_hyb}, with the dwell-times reported in Table \ref{tb:Boiler_par_3}.
\begin{table}[h!]
	\begin{center}
		\caption{Hybrid automaton dwell times (in HL steps $T_{\scriptscriptstyle\rm H}$).}
		\begin{tabular}{l||c|c|c} 
			\textbf{Boiler} & {$\chi_{\scriptscriptstyle \rm{OFF}\rightarrow \rm{ST}}$} &  $\chi_{\scriptscriptstyle \rm{ST}\rightarrow \rm{ON}}$ 			& $\chi_{\scriptscriptstyle \rm{ON}\rightarrow \rm{OFF}}$\\			
			\hline
			$i$ & $2$ & $2$ & $3$\\
			\hline
		\end{tabular}
	\label{tb:Boiler_par_3}
	\end{center}
\end{table}
In this case-study all generators have the same transition times.\\
It is worth emphasizing that, as reported in Figure \ref{fig:SG_Ens}, the reference trajectory is naturally given in terms of steam demand $\bar{q}_{\scriptscriptstyle\rm s}^{\scriptscriptstyle\rm Dem}$ and converted into equivalent gas target using the static gain of the ensemble $\bar{q}_{\scriptscriptstyle\rm g}^{\scriptscriptstyle\rm Dem} = \bar{g} \bar{q}_{\scriptscriptstyle\rm s}^{\scriptscriptstyle\rm Dem} + \bar{\gamma}$. In particular, the  reference target for the fuel flow-rate of the ensemble incorporates only the units in mode ON. While the high-level optimizer considers the consumption and the relative costs of the steam generators also in startup modes, we recall that the ensemble model considers just the producing boiler, i.e., in mode ON. The scope of the MPC layer is indeed a robust reference tracking for the ensemble and not an economic optimization, which is the target of the high-level optimization.\\
As discussed in Section \ref{subsec:Ref_model}, a requirement is that all the reference models share the same dynamic and output matrices $\hat{A}_i$ and $\hat{C}_i$, respectively. Conceptually, they can be arbitrarily chosen by the designer, e.g., by imposing a desired dynamic matrix or an ''averaged'' one for all the subsystems of the ensemble. In this work, we select a specific unit as the reference dynamic model: therefore, we impose the matrices of the first steam generator for the reference model, i.e.  $\hat{A}_i=A_1$ and $\hat{C}_i=C_1$. As a consequence, the state reduction map is simply $\beta_{\scriptscriptstyle i} = I_{\scriptscriptstyle n}$ for all the subsystems and gain consistency conditions reduce to \eqref{eq:gain_cons_both}.\\
In Figure \ref{fig:ref_models}, the comparison of the step response of each system $\mathcal{L}_{i}^{\scriptscriptstyle \rm{CL}}$ with its reference model $\hat{\mathcal{L}}_{i}$ is shown: the gain consistency conditions  \eqref{eq:gain_cons_both} guarantee that at steady state the actual and reference models reach the same value.\\
The maximum amplitude of the disturbance $\bar{w}$ in the ensemble model is evaluated by imposing the maximum variation of the input equal to  $\Delta \bar{u} =  0.4 [kg/s]$, resulting in $\|{ \bar{w}} \|_{\infty} \leq 1\times 10^{-3} [kg/s]$.\\
\begin{table}[b]
	\begin{center}
		\caption{Control architectures features for performance comparison.}
		\begin{tabular}{l||l|l|c|c} %
			\textbf{Strategy} & \textbf{HL} &  \textbf{ML-MPC} 	& \textbf{LL}  & \textbf{Test}	\\			
			\hline
			\textbf{HL OPT} & $\alpha_{\scriptscriptstyle i} \leftarrow$\eqref{eq:opt_simpl} ($l_{\scriptscriptstyle i}$ eq.\eqref{eq:alpha_opt_descr})& Ensemble & PI & 1\&2 \\
			\textbf{NO HL} & $\alpha_{\scriptscriptstyle i} \leftarrow 1/{\sum x_{\scriptscriptstyle{\rm ON}\, i}}$ & Ensemble & PI & 1\&2 \\
			\textbf{HL $\eta$-OPT } & $\alpha_{\scriptscriptstyle i} \leftarrow$\eqref{eq:opt_simpl} ($l_{\scriptscriptstyle i}=-\eta_{\scriptscriptstyle i}x_{\scriptscriptstyle{\rm ON}\, i}$)& Ensemble & PI &  1\&2 \\
			\textbf{C-MPC} & - & Centralized & PI & 2 \\
			\hline
			\multicolumn{5}{l}{Test 1: Unit 3 with scheduled maintenance, piece-wise demand}\\
			\multicolumn{5}{l}{Test 2: All units available, noisy demand}
		\end{tabular}
		\label{tb:cs}
	\end{center}
\end{table}
We compare, in simulation, the performance of the proposed control scheme (HL OPT) with two alternative ones, obtainable with different strategies, see Table \ref{tb:cs}: NO HL, where the sharing factors are not optimized during operation, but predefined and fixed (e.g., by equally splitting the load on available units,  $\alpha_{\scriptscriptstyle i} = 1/\sum x_{\scriptscriptstyle{\rm ON}\, i}$), and HL $\eta$-OPT, where the units are activated in a round-robin fashion according to their efficiency ranking. For all these schemes, the robust MPC, see \eqref{eq:MPC_Rob1}, is synthesized on the ensemble configuration defined at HL, 
	using the {steady-state} terminal condition,  with a prediction horizon $N_{\scriptscriptstyle\rm M} = 10$. 
	The constraints, imposed according to the tube-based paradigm, are enforced in a tightened way to the unperturbed system variables.\\ 
	The tracking performances are also compared with the ones of a centralized MPC scheme (C-MPC), which controls directly all the subsystem inputs, $u_{\scriptscriptstyle i}$. Two scenarios are proposed: \emph{Test 1} considers a maintenance schedule for Boiler 3, with a piece-wise constant demand; \emph{Test 2} shows the behavior of the schemes with a noisy demand, to assess the operational cost and the tracking performance.
\subsubsection{Test 1}
\begin{figure}[t]
	\centering
	\includegraphics[width=1\linewidth]{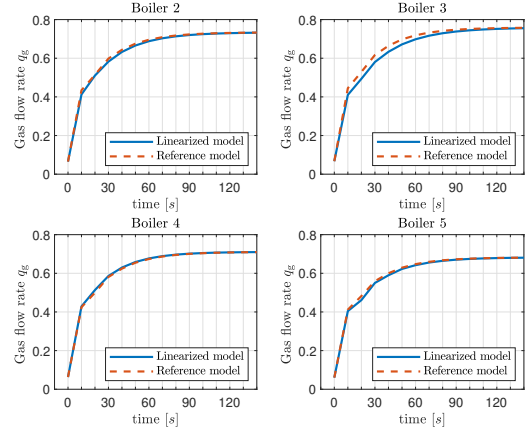}
	\caption{Step-response of the actual (solid line) and reference  models (dashed line).}
	\label{fig:ref_models}
\end{figure}%
\begin{figure}[tbh]
	\centering
	\includegraphics[width=1.0\linewidth]{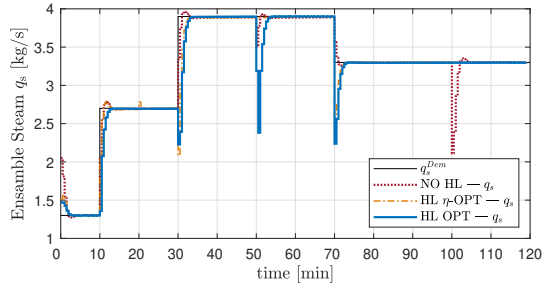}
	
	\caption{Steam demand for the ensemble $\bar{q}_{\scriptscriptstyle\rm s}^{\scriptscriptstyle\rm Dem}$ (black) tracked by Ensemble-MPC at ML, with HL OPT strategy (solid blue), HL $\eta$-OPT (dot-dashed orange), and NO HL (dotted red).}
	\label{fig:Q}
\end{figure}%
\begin{figure}[tb]
	\centering
	\includegraphics[width=1\linewidth]{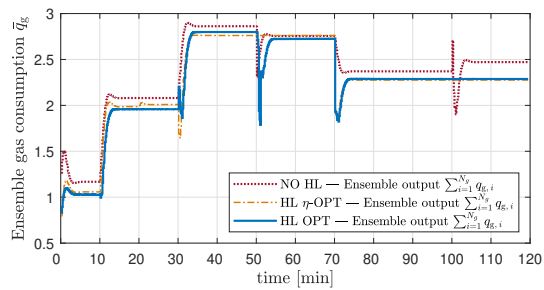}
	\caption{Ensemble gas consumption. The reference target depends on the ensemble configuration, since different sharing factors change the overall gain.}
	\label{fig:Qg}
\end{figure}%
We assume Boiler $3$ to be unavailable in the time range $t=[50,80)$ min and we compare the behavior of the proposed scheme with HL $\eta$-OPT and with NO HL. The idea here is to focus on the role of the HL control layer on the overall performances. Figure \ref{fig:Q} shows the tracking of steam demand with the three considered strategies. Note that the overshoots are due to the plug/unplug operations.
\begin{figure}[tb]
	\centering
	\includegraphics[width=1\linewidth]{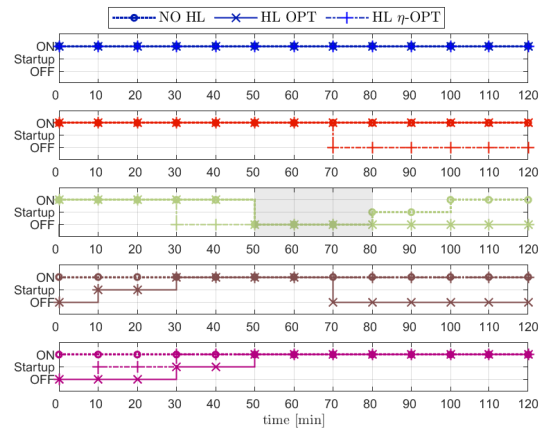}
	\caption{Operating mode of each subsystem. Boiler~$3$ temporal unavailability shown by gray region.}
	\label{fig:delta}
\end{figure}%
Figure \ref{fig:Qg} shows the reference tracking performances of the natural gas signal. 
Due to the unequal overall gain of the ensemble in the three alternative configurations, natural gas trajectories are different. This is more evident in the period $t=[30, 70)$ min even if the steam demand is the same. This is related to a difference in the subsystem's efficiency. Recall that the gas consumption is given by \eqref{eq:IN-OUT-B1} for each boiler  and the ensemble efficiency is given by the $\alpha$-weighted combination of such equations.\\ 
In Figure \ref{fig:delta}, the operating modes of each unit are shown: at $t=50$ min, Boiler $3$ is forced to OFF mode, for prescribed unavailability (e.g., for  maintenance reasons) shown by a gray area.  With HL optimization, Boiler $4$ is activated in place of Boiler $3$. Note that, even if Boiler $5$ has a higher efficiency with respect to Boiler $4$, the latter is chosen in the first place due to its lower start-up costs $\lambda_{\scriptscriptstyle{\rm{ST}} \,\scriptstyle i}$; with HL $\eta$-OPT, instead, the different efficiency-based criterion for boiler activation leading to slightly larger overall costs.\\
As shown in Figure \ref{fig:alpha}, when the global steam demand rises, new generators are added to the ensemble based, not only on the subsystem efficiency rank, but also on the associated operating costs $\lambda_{\scriptscriptstyle{\rm{ON}} \,\scriptscriptstyle i}$ and  $\lambda_{\scriptscriptstyle{\rm{ST}} \,\scriptscriptstyle i}$, which are different for each generator.  In the NO HL scenario, the weights $\alpha_i$ are adjusted only to consider that just four generators are available.
\begin{figure}[tb]
	\centering
	\includegraphics[width=1\linewidth]{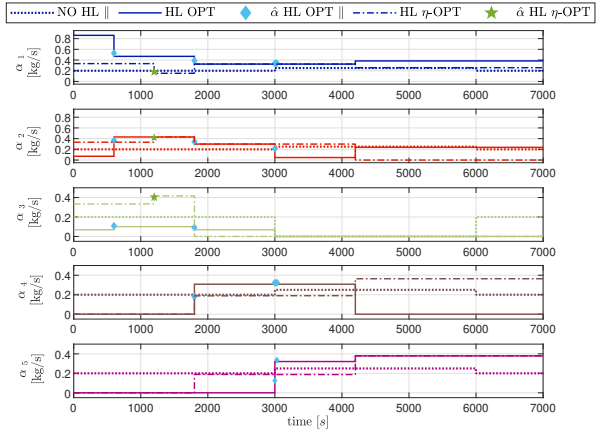}
	\caption{High-level sharing factors with HL OPT strategy (solid, transient $\hat{\alpha}$ diamonds), HL $\eta$-OPT (dot-dashed, transient $\hat{\alpha}$ stars), and NO HL (dotted).  }
	\label{fig:alpha}
\end{figure}%
 When the transition is sharp, the abrupt change of $\alpha_i$ could lead one of the subsystems out of its local ranges. If so, the MPC optimization problem may become infeasible. In response to that, the control architecture will compute a transient solution by considering the sharing factors as an additional set of optimization variables, as discussed in Section \ref{subsec:Trans_MPC}: the nonlinear program provides the closest feasible configuration  to the target computed at the top-level. In Figure \ref{fig:alpha}, the  sharing factors computed at high-level, with the three strategies are shown by different line styles. When the high-level solution is reachable in one medium-level step, the optimal and actual points coincide and just target is shown,
otherwise the ensemble is guided to the high-level optimal configuration by a smooth shift through temporary configurations (a diamond for the HL OPT and a star for HL $\eta$-OPT), computed solving the NLP. Figure \ref{fig:qg_i} shows that also local constraints are respected.\\ %
\begin{figure}[tb]
	\centering
	\includegraphics[width=1\linewidth]{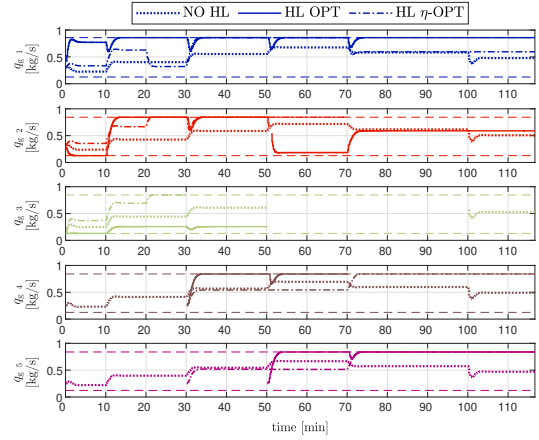}
\caption{In each subplot, the gas flow-rate $q_{\rm g, \, \it{i}}$ of the units.}
	\label{fig:qg_i}
\end{figure}%
The simulation is executed in Matlab on a Intel$^{\text{\textregistered}}$Core$^{\text{\texttrademark}}$~i7-8550U CPU~1.80GHz, RAM 16 GB, with \texttt{SCIP} solver \cite{GamrathEtal2020OO} for HL optimization and transitional configurations, and \texttt{quadprog} for medim-level QP. The HL optimization takes an average time of $3.28$s ($\pm 0.53$s), while the ML QP takes $0.13$s ($\pm 0.02$s), where the RPI computation requires $0.08$s. Instead, the NLP for transitional configuration requires up to $40$s. %
\subsubsection{Test 2}
A second test is done to compare the performance of the proposed scheme and to demonstrate also the robustness of the control architecture in presence of possibly significant errors on the demand forecast. Here the focus is both on the HL, considering the operational cost, and on the ML, by measuring the tracking performance. The latter is assessed by considering a further alternative scheme consisting of a centralized MPC, which governs directly the input $u_{\scriptscriptstyle i}$ of each subsystem. It is worth noting, that this controller cannot manage the HL dynamics related to mode transitions, i.e., start-ups, and plug-and-play operations, thus for this case it does not really make sense to quantify the related HL operational cost. However, given a fixed number of active generators, this represents the best tracking controller. The performance metric is given by $J^{{\scriptscriptstyle\rm tr}}_{\scriptscriptstyle\rm M} = \sum_{k} \|y(k)-r\|^{\scriptscriptstyle 2}$. Instead, the operational cost $J^{{\scriptscriptstyle\rm op}}$, is computed as \eqref{eq:alpha_opt_descr}.\\
 The disturbed demand is given by $r\left(k\right)=\bar{q}_s^{\rm\scriptscriptstyle Dem}\left(\left\lfloor h/\mu\right\rfloor\right)+v\left(k\right)$ with the noise term $v\left(k\right)\sim n\left(0,\ \sigma\right)$, with $\sigma={1.25\%\bar{q}}_s^{\rm\scriptscriptstyle Dem}$. 
In addition, at $t=90$ (resp. $100$) min a downward (resp. upward) step disturbance is given, with $d={\pm4\%\bar{q}}_s^{\rm\scriptscriptstyle Dem}$ thus with the noise term\footnote{Note that, merely for clarity of the resulting plots, we have reduced the high-frequency component of the noise by setting a lower standard deviation.} $v\left(k\right)\sim n\left(d,\ \sigma/10\right)$. \\		
\begin{table}[h!]
	\begin{center}
		\caption{Operational cost, $J^{\rm\scriptscriptstyle op}$,  (scaled on NO HL cost) - and tracking cost, $J^{\rm\scriptscriptstyle tr}$,  (scaled on C-MPC cost).}
		\begin{tabular}{l||c|c|c|c} 
			\textbf{Cost} & {NO HL} &  HL OPT 	& HL $\eta$ OPT & C-MPC	\\			
			\hline
			${J^{\rm\scriptscriptstyle op}/J^{\rm\scriptscriptstyle op}_{\rm\scriptscriptstyle NO\, HL}} [-]$  & $1.00$ & $0.78$  & $0.80$ & - \\
			${J^{\rm\scriptscriptstyle tr}/J^{\rm\scriptscriptstyle tr}_{\rm\scriptscriptstyle C-MPC}} [-]$  & $2.30$ & $3.09$  & $3.48$ & $1.00$ \\
			\hline
		\end{tabular}
		\label{tb:cost}
	\end{center}
\end{table}
In Figure \ref{fig:Qg_multi}, the tracking of the natural gas for the ensemble is reported for the different control strategies. Note that the different overall efficiency of the ensemble leads to distinct gas flow-rates, even if the steam demand is the same, see Figure \ref{fig:Qs_multi}. At medium level this demand is disturbed by an additive noise term, the mismatch between the piece-wise reference is managed by the MPC: as the reference $\hat{r}$ is a decision variable at ML, this MPC formulation can deal also with infeasible references.
Typically, an increased demand might become unreachable, while  lower actual demand can be easily managed: in $t=[90,\ 100)$ min, with the downward step disturbance, ML can track the actual demand by keeping the same sharing factors. Instead in $t=[100,\ 110)$ min, the global generation cannot reach the actual target. However, the controller robustly gives a feasible solution, which minimizes the distance from the target.
The event-based optimization of the HL sharing factors is applied at $t=102.5$ min, when a bias on the demand is detected: a new HL optimization is triggered on an updated demand forecast, which includes the bias. The sharing factors are adapted  to achieve the increased demand.	
The best tracking performance is obtained by C-MPC, that does not operate on an overall ensemble model, but controls directly each subsystem. However, it does not provide flexibility to system changes and scalability properties.\\
A good tracking performance is apparently given by the NO HL scheme, where demand at ML is tracked by an ensemble-MPC, with all the units sharing equally the load. The controller at ML is the same used for HL $\eta$-OPT and HL OPT, where the plug-\&-play operations negatively impacts the tracking. Note that, instead, the overall operational cost $J^{\rm\scriptscriptstyle op}$ is better in case the HL optimization is performed.
The operational and tracking costs of the four strategies are compared in Table \ref{tb:cost}. 
\begin{figure}[tb]
	\centering
	\includegraphics[width=1\linewidth]{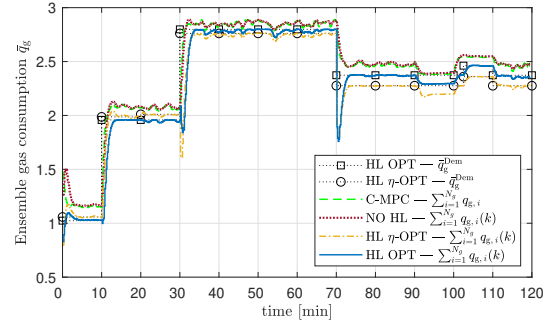}
	\caption{Natural gas consumption of the ensemble. Comparison of the four strategies.}
	\label{fig:Qg_multi}
\end{figure}%
\begin{figure}[tb]
	\centering
	\includegraphics[width=1\linewidth]{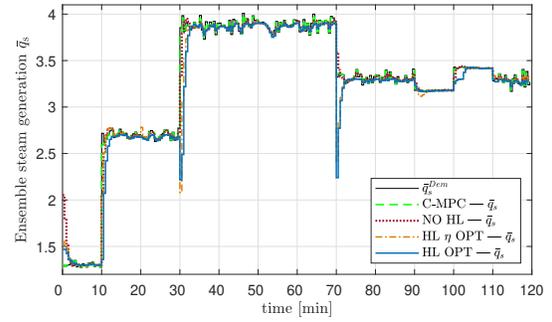}
	\caption{Ensemble steam flow-rate. Comparison of the four strategies. }
	\label{fig:Qs_multi}
\end{figure}%
Regarding the computational complexity, the proposed method maintains constant the dimension of the QP problem to be solved at medium level, by relying on the ensemble model. On the contrary, the dimension of the QP with the C-MPC grows linearly with respect to the number of considered units, see Table \ref{tb:cpu}. 
A proportional dimension increase affects also the high-level MIP: in this case the computational impact is greater. Note that it still within the HL sampling time, $T_{\rm\scriptscriptstyle H} = 10$ minutes. Note that, however, the HL should run offline, and so its computational complexity does not impact on the real-time feasibility of the hierarchical scheme.
\begin{table}[h]
	\begin{center}
		\caption{Computational complexity.}
		\begin{tabular}{l l||c|c|c} 
			&\textbf{No. of units} & $5$ &  $10$ 	& $15$\\ 			
			\hline
			\multirow{4}{*}{E-MPC}& [vars.] & $14$ & $14$ & $14$ \\ 
			&CPU [s] mean & $0.158$ & $0.161$ & $0.131$ \\ %
			&CPU [s] min & $0.106$ & $0.102$ & $0.023$ \\ %
			&CPU [s] max & $0.271$ & $0.463$ & $0.333$ \\ %
			\hline
			\multirow{4}{*}{C-MPC}& [vars.] & $51$ & $101$ & $151$ \\ 
			&CPU [s] mean & $0.141$ & $0.170$ & $0.205$ \\ %
			&CPU [s] min & $0.102$ & $0.127$ & $0.162$ \\ %
			&CPU [s] max & $0.808$ & $0.819$ & $0.876$ \\ %
						\hline
						\multirow{2}{*}{HL OPT}& [vars.] & $0.5$k & $1.1$k & $1.5$k \\ %
						&CPU [s] mean & $3.28$ & $6.54$ & $21.44$ \\ %
						\hline
		\end{tabular}
		\label{tb:cpu}
	\end{center}
\end{table}	

\section{Conclusions}
In this paper a hierarchical control scheme has been proposed for the coordination of an ensemble of steam generators, which must cooperate to fulfill a common load. The definition of an ensemble reference model, as proposed here, permits to solve the medium level tracking MPC in a scalable and flexible way, as its dimension does not grow with the number of steam generators in the ensemble. Thanks to the model reformulation, the ensemble model can be simply obtained from the high level and updated online. The model configuration is determined by the high-level mixed-integer optimization that computes the optimal number of generators to be included in the ensemble and their shares of steam production by minimizing the operating cost and considering global and subsystem constraints.\\ 
The accuracy of demand forecast  impacts the solution quality: generally, forecast mismatch is managed at medium level, with a small degradation of the overall cost. However, if units are committed with a greedy policy with active ones working already at maximum, any higher actual demand cannot be fully sustained, as an additional boiler would be needed, but the start-up dynamics might impede it.
This can be managed by  tightening  the subsystem input/output constraints, at HL level, to prevent such condition, even if the feasibility at medium level is guaranteed by the presence of the reference point among the decision variables. How to properly set	this tightening will be studied.
Future work will consider the improvement of the multi-layer scheme by comparing the overall performance with the implementation of an additional low-level shrinking MPC control to further address the local model mismatch. We also envision to solve the high level optimization in a distributed framework.

\bibliographystyle{IEEEtran}
\bibliography{references}             

\vskip -2.2\baselineskip plus -1fil
\begin{IEEEbiographynophoto}{Stefano Spinelli} received the master's degree in aeronautical engineering and the Ph.D. degree in Information Technology from the Politecnico di Milano, in 2012 and 2021, respectively. Since 2013 he is a researcher at Consiglio Nazionale delle Ricerche at the Istituto di Sistemi e 	Tecnologie Industriali Intelligenti per il Manifatturiero Avanzato. His research interests include optimization and control of large-scale systems, and energy-aware scheduling of industrial processes for demand response.
\end{IEEEbiographynophoto}%
\vskip -2.2\baselineskip plus -1fil
\begin{IEEEbiographynophoto}{Marcello Farina} received the master’s degree in electronic engineering and the Ph.D. degree in information engineering from the Politecnico di Milano, Milan, Italy, in 2003 and 2007, respectively. He was Visiting Student with the Institute for Systems Theory and Automatic Control, Stuttgart, Germany, in 2005. He is currently an Associate Professor with the Dipartimento di Elettronica, Informazione, e Bioingegneria, Politecnico di Milano. His current research interests include distributed, hierarchical and decentralized state estimation and control, stochastic model predictive control, and applications, including mobile robots, sensor networks, and energy supply systems.	
\end{IEEEbiographynophoto}%
\vskip -2.2\baselineskip plus -1fil
\begin{IEEEbiographynophoto}{Andrea Ballarino} received the master’s degree in Computer Science and Engineering - with specialisation in  robotics and industrial automation - in 2001, from the Politecnico di Milano. He is currently responsible of 2MaCS - Machine and Manufacturing Control System Group within the Istituto di Sistemi e Tecnologie Industriali Intelligenti per il Manifatturiero Avanzato of Consiglio Nazionale delle Ricerche. His research interests focus on methodologies and tools for distributed automation systems design and development, adaptive and sustainable automation systems, model predictive control and real time process optimization, hybrid control systems. 
\end{IEEEbiographynophoto}%
\end{document}